\documentclass[11pt,a4paper]{article}
\usepackage{graphicx}
\usepackage{amssymb}
\usepackage{amsmath}
\usepackage{amsfonts}
\usepackage{dsfont}
\usepackage{mathtools}
\usepackage{array}
\usepackage{soul}
\usepackage{rotating}
\usepackage{bbold,amsfonts}
\allowdisplaybreaks

\usepackage[utf8]{inputenc}
\usepackage{bm}
\usepackage{xcolor}
\usepackage{float}
\usepackage{braket}
\usepackage{placeins}
\usepackage[height=8.8in,width=6.45in]{geometry}
\usepackage[font=small,labelfont=bf]{caption}
\usepackage[hidelinks]{hyperref}
\usepackage{booktabs,float,slashed}
\usepackage{standalone}
\usepackage{caption}
\usepackage{subcaption}
\usepackage{makecell}
\usepackage[maxbibnames=99,sorting=none,giveninits=true,backend=biber,style=numeric-comp,sortcites,doi=false,hyperref=true]{biblatex}
\renewbibmacro{in:}{}
\usepackage{hyperref}
\DeclareFieldFormat{doilink}{\iffieldundef{doi}{#1}{\href{https://doi.org/\thefield{doi}}{#1}}}
\DeclareFieldFormat[article,periodical]{volume}{\mkbibbold{#1}}
\DeclareFieldFormat[article,periodical]{journaltitle}{#1}
\DeclareFieldFormat[article,periodical]{pages}{#1}

\usepackage{xpatch}
\xpatchbibdriver{article}
  {\usebibmacro{journal+issuetitle}%
   \newunit
   \usebibmacro{byeditor+others}%
   \newunit
   \usebibmacro{note+pages}}
  {\printtext[doilink]{%
     \usebibmacro{journal+issuetitle}%
     \newunit
     \usebibmacro{byeditor+others}%
     \newunit
     \usebibmacro{note+pages}}}
  {}{}
\numberwithin{equation}{section}
\newcommand{\beq}{\begin{equation}}
\newcommand{\eeq}{\end{equation}}

\newcommand{\ii}{\mathrm{i}}

\makeatletter
\newcommand*{\letterdef@}{}
\newcommand*{\letterdef}[3]{%
	\def\letterdef@##1{\expandafter\newcommand\csname #1\endcsname{#2{##1}}}%
	\@tfor\@tempa :=#3\do{\expandafter\letterdef@\expandafter{\@tempa}}}
\makeatother
\letterdef{c#1} {\mathcal}{ABCDEFGHIJKLMNOPQRSTUVWXYZ} % \cX = \mathcal{X}
\letterdef{rm#1}{\mathrm} {dDeimM} % \rmX = \mathrm{X} for X in {dDmM}

\title{Draft integrated correlators and quivers}

\begin{document}

\begin{titlepage}

\begin{flushright}
%\footnotesize
\small
\texttt{HU-EP-24/11}
\end{flushright}

\vspace*{10mm}
\begin{center}
{\LARGE \bf 
Integrated correlators at strong coupling 

\vspace{0.2cm}

in an orbifold of $\mathcal{N}=4$ SYM 
}

\vspace*{15mm}

{\Large  Alessandro Pini${}^{\,a}$ and Paolo Vallarino${}^{\,b,c}$}

\vspace*{8mm}

${}^a$ Institut f{\"u}r Physik, Humboldt-Universit{\"a}t zu Berlin,\\
     IRIS Geb{\"a}ude, Zum Großen Windkanal 2, 12489 Berlin, Germany  
     \vskip 0.3cm

${}^b$ Universit\`a di Torino, Dipartimento di Fisica,\\
			Via P. Giuria 1, I-10125 Torino, Italy
			\vskip 0.3cm
			
${}^c$   I.N.F.N. - sezione di Torino,\\
			Via P. Giuria 1, I-10125 Torino, Italy

\vskip 0.8cm
	{\small
		E-mail:
		\texttt{alessandro.pini@physik.hu-berlin.de;paolo.vallarino@unito.it}
	}
\vspace*{0.8cm}
\end{center}

\begin{abstract}
We consider the $4d$ $\mathcal{N}=2$ superconformal quiver gauge theory obtained by a $\mathbb{Z}_2$ orbifold of $\mathcal{N}=4$ super Yang-Mills (SYM). By exploiting supersymmetric localization, we study the integrated correlator of two Coulomb branch and two moment map operators  and the integrated correlator of four moment map operators, determining exact expressions valid for any value of the 't Hooft coupling in the planar limit. Additionally, for the second correlator, we obtain an exact expression also for the next-to-planar contribution. Then, we derive the leading terms of their strong-coupling expansions and outline the differences with respect to the $\mathcal{N}=4$ SYM theory.

\end{abstract}
\end{titlepage}
\setcounter{tocdepth}{2}
\tableofcontents

\section{Introduction}

Four dimensional $\mathcal{N}=2$  superconformal  field theories (SCFTs) are a very worthwhile framework to examine the strong-coupling limit of interacting quantum field theories. In general, extended supersymmetry together with conformal invariance allows to employ many powerful techniques, such as integrability, conformal bootstrap, supersymmetric localization \cite{Pestun:2007rz} and the AdS/CFT correspondence \cite{Maldacena:1997re}, that can be exploited to evaluate various  observables  in the strong-coupling regime of these theories.

In this work we will focus on correlation functions of scalar superconformal primary operators. In particular we will consider operators that are the top components of two different short multiplets of the $\mathfrak{su}(2,2|2)$ superconformal algebra. The first ones are the Coulomb branch operators $\mathcal{O}_k$. They belong to the $\mathcal{E}_k$ short multiplet\footnote{We follow the conventions of \cite{Dolan:2002zh} for denoting the multiplets of the $\mathfrak{su}(2,2|2)$ algebra.},
are annihilated by all the supercharges with a given chirality and are chiral primary operators (CPOs). 
This implies that the  $\mathcal{O}_k$ are labelled by their integer conformal dimension $k$, which is a protected quantity.  

The other superconformal primary operators belong to the $\mathcal{\hat{B}}_R$ short multiplet of $\mathfrak{su}(2,2|2)$. These are Higgs branch operators which are annihilated by supercharges with opposite chirality and have a protected conformal dimension that is just a function of the charge under the $SU(2)_R$ R-symmetry subgroup. We just consider correlation functions involving Higgs branch operators of conformal dimension 2, which are the top component of the short multiplet  $\mathcal{\hat{B}}_1$ and are constructed as bilinears of the scalar fields in the hypermultiplets. We refer to these operators as moment map operators and we denote them by $\mathcal{J}$. 

A powerful technique which is very efficient in computing correlation functions of superprimary operators in $\mathcal{N}=2$ supersymmetric gauge theories is localization,  that maps the computation of path integrals to finite-dimensional integrals over matrix elements (see \cite{Pestun:2016zxk}  for a review). The basic idea to evaluate correlators of local operators consists in considering  deformations of the original gauge theory preserving $\mathcal{N}=2$ supersymmetry, placing the deformed theory on the sphere $\mathbb{S}^4$ and then applying supersymmetric localization to compute the deformed partition function  \cite{Pestun:2007rz}. In order to better understand this point, it is important to recall that in any $\mathcal{N}=2$ SCFT there are two possible deformations that preserve $\mathcal{N}=2$ Poincaré supersymmetry \cite{Argyres:2015ffa,Cordova:2016xhm}: deformations by an integral of a Coulomb branch operator $\mathcal{O}_k$ through a coupling $\tau_k$ and mass $m$ deformations proportional to an integral of the aforementioned moment map operators. Thus it appears clear that by taking appropriate derivatives of the deformed partition function with respect to $\tau_k$ and $m$, we obtain integrated correlators of CPOs and moment map operators.

This procedure was first applied in \cite{Gerchkovitz:2016gxx} to extremal correlators of Coulomb branch operators, i.e. correlation functions of $n-1$ chiral operators and one anti-chiral one. In this specific case, however, exploiting superconformal symmetry it was proven that this method allows to fully determine the functional form of the correlators, namely to evaluate un-integrated extremal correlators of CPOs. Subsequently, following this approach, many results for these observables have been obtained in different examples of $4d$ $\mathcal{N}=2$ SCFTs, either at large or finite $N$ \cite{Baggio:2014sna,Baggio:2015vxa,Baggio:2016skg,Rodriguez-Gomez:2016cem,Rodriguez-Gomez:2016ijh,Pini:2017ouj,Billo:2017glv,Beccaria:2020hgy,Galvagno:2020cgq,Beccaria:2021hvt,Fiol:2021icm,Billo:2021rdb,Billo:2022xas,Fiol:2022vvv,Billo:2022gmq,Billo:2022fnb,Beccaria:2022ypy,Bobev:2022grf,Billo:2022lrv,Preti:2022inu,Nunez:2023loo}. 

More recently, the above procedure has been employed to compute a new class of 4-point functions \cite{Binder:2019jwn}, namely 4-point correlators integrated with a measure completely fixed by superconformal symmetry. In particular it was proven that by taking four derivatives of the partition function on $\mathbb{S}^4$ of the mass deformation of $\mathcal{N}=4$ SYM, called $\mathcal{N}=2^*$ theory \cite{Pestun:2007rz,Russo:2013kea}, one can derive the integrated correlator of four superconformal primaries in $\mathcal{N}=4$ SYM, precisely
\begin{align}
\label{OOJJ}
\partial_{\tau_p}\partial_{\overline{\tau}_p}\partial_{m}^2 \log\mathcal{Z}_{\mathcal{N}=2^{*}}\Big\vert_{m=0} = \int \prod_{i=1}^{4}dx_i  \, \mu(\{x_i\}) \, \langle \mathcal{O}_p(x_1)\mathcal{\overline{O}}_p(x_2)\mathcal{J}(x_3)\mathcal{J}(x_4) \rangle_{\mathcal{N}=4} \,.
\end{align}
In writing \eqref{OOJJ} we have thought the $\mathcal{N}=4$ SYM in a $\mathcal{N}=2$ language, i.e. as a $\mathcal{N}=2$ SCFT with a massless hypermultiplet in the adjoint representation of the gauge group. In this sense $\mathcal{O}_p(x_i)$ and $\mathcal{\overline{O}}_p(x_i)$ are the chiral and anti-chiral Coulomb branch operators, whereas $\mathcal{J}(x_i)$ are the Higgs branch moment map operators. The explicit expression of the integration measure $\mu(\{x_i\})$ can be found in \cite{Binder:2019jwn}. 

A different integrated correlator was identified in \cite{Chester:2020dja}
\begin{align}
\label{4mintcorr}
\partial_{m}^4 \log\mathcal{Z}_{\mathcal{N}=2^{*}}\Big\vert_{m=0} = \int \prod_{i=1}^{4}dx_i  \, \hat{\mu} (\{x_i\}) \, \langle \mathcal{J}(x_1)\mathcal{J}(x_2) \mathcal{J}(x_3)\mathcal{J}(x_4) \rangle_{\mathcal{N}=4} \,,
\end{align}
where $\hat{\mu}(\{x_i\})$ is a distinct integration measure.

It is worth emphasising that, even though we have now distinguished the superprimary operators between Coulomb branch operators and Higgs branch operators, in $\mathcal{N}=4$ SYM, if we consider $p=2$ in \eqref{OOJJ}, both $\mathcal{O}_2$ and $\mathcal{J}$ actually belong to the stress-tensor multiplet, transforming in the $\textbf{20}'$ representation of the R-symmetry group $SU(4)$. Remarkably, this means that, in that case, \eqref{OOJJ} and \eqref{4mintcorr} give information about the same correlation function of superprimary operators.

These important results have been further investigated in \cite{Chester:2019pvm,Chester:2019jas,Chester:2020vyz,Green:2020eyj,Dorigoni:2021bvj,Dorigoni:2021guq,Dorigoni:2021rdo,Alday:2021vfb,Dorigoni:2022cua,Dorigoni:2022zcr,Collier:2022emf,Wen:2022oky,Paul:2022piq,Paul:2023rka,Brown:2023cpz,Brown:2023why,Pufu:2023vwo,Brown:2023zbr,Dorigoni:2023ezg,Billo:2023ncz,Alday:2023pet,Brown:2024tru}, where many new features have been studied, such as the large-$N$ expansion up to very high orders  using topological recursion, the modular and exact properties of these quantities, the extension to all simple gauge groups and to more general $\frac{1}{2}$-BPS operators, the large-charge limit, integrated correlators with the insertion of determinant operators and also mixed
integrated correlators involving two local operators and a Wilson line . All these results concern the $\mathcal{N}=4$ SYM theory.

Despite the huge amount of results found for the $\mathcal{N}=4$ SYM, less attention has been devolved to the study of the above mentioned integrated correlators in purely $\mathcal{N}=2$ theories.
In particular, in \cite{Chester:2022sqb} a relation analogous to \eqref{4mintcorr} between four mass derivatives of the mass-deformed free energy of a generic $4d$ $\mathcal{N}=2$ SCFT on $\mathbb{S}^4$ and the integrated correlator of four moment map operators was derived and analysed in detail for $\mathcal{N}=2$ superconformal QCD (SQCD) with $SU(2)$ gauge group. Subsequently, in \cite{Fiol:2023cml} the same correlator was examined in the large-$N$ limit of $SU(N)$ SQCD. In \cite{Behan:2023fqq} the integrated 4-point function of these Higgs branch operators and its holographic dual in terms of gluons scattering were investigated in a $4d$ SCFT with $USp(2N)$ gauge group, four hypermultiplets in the fundamental representation, one in the antisymmetric and $SO(8)$ flavour symmetry. Finally, in \cite{Billo:2023kak} the integrated correlator involving two Coulomb branch operators and two moment map operators, analogous to \eqref{OOJJ}, was considered in the large-$N$ limit of a $\mathcal{N}=2$ SCFT with $SU(N)$ gauge group, the so-called \textbf{E}-theory, that has one symmetric and one antisymmetric hypermultiplet.

In this article we study these integrated correlators in the $\mathcal{N}=2$ SCFT that is obtained from a $\mathbb{Z}_2$ orbifold projection of $\mathcal{N}=4$ SYM. This theory has a quiver-like structure with 2 nodes: it is made up of $SU(N)\times SU(N)$ gauge group with two $\mathcal{N}=2$ hypermultiplets in the  bifundamental representation of the gauge groups. For simplicity, we just consider the orbifold fixed point of the theory where the two Yang-Mills couplings are equal. This is the most symmetric configuration and possesses a known holographic dual given by Type II B string theory on a space-time of the type $AdS_5\times \mathbb{S}^5/\mathbb{Z}_2$ \cite{Kachru:1998ys,Gukov:1998kk}.
Over the last few years, many BPS-observables in the large-$N$ limit of this superconformal theory have been examined, both at weak and strong-coupling, exploting supersymmetric localization techniques, such as the free energy and the VEV of one or more circular Wilson loops \cite{Rey:2010ry,Zarembo:2020tpf,Fiol:2020ojn,Ouyang:2020hwd,Beccaria:2021ksw,Galvagno:2021bbj,Pini:2023lyo,Beccaria:2023kbl} and the 2- and 3- point functions of chiral/anti-chiral primary operators \cite{Pini:2017ouj,Galvagno:2020cgq,Fiol:2021icm,Billo:2021rdb,Billo:2022gmq,Billo:2022fnb,Billo:2022lrv,Beccaria:2022ypy}.

The goal of this work is twofold. First, we concentrate on the integrated four-point function of two Coulomb branch operators and two moment map operators. Therefore, we focus on the quantity
\begin{align}
\label{Z2intcorr}
\mathcal{C}_{f,p} \equiv \partial_{\tau_p}\partial_{\overline{\tau}_p}\partial_{m_{f}}^2 \log\mathcal{Z}(m_1,m_2)\Big\vert_{m_1=m_2=0}\,,
\end{align}
where $f=1,2$ and $\mathcal{Z}(m_1,m_2)$ is the mass-deformed partition function of the $\mathbb{Z}_2$ quiver gauge theory. In this quiver, however, we can have two different kinds of deformations involving the chiral primary operators. Indeed we can introduce untwisted, $\mathcal{O}_p^+$, and twisted, $\mathcal{O}_p^-$, CPOs under the $\mathbb{Z}_2$ action that exchanges the two gauge groups \cite{Billo:2021rdb,Billo:2022fnb}
\begin{align}
\label{utCPO}
\mathcal{O}^+_p(x)=\frac{1}{\sqrt{2}} \biggl[ \text{tr}\phi_0(x)^p + \text{tr}\phi_1(x)^p \biggr]\,,\ \ \ \ \ \mathcal{O}^-_p(x)=\frac{1}{\sqrt{2}} \biggl[ \text{tr}\phi_0(x)^p - \text{tr}\phi_1(x)^p \biggr]\,,
\end{align}
where $\phi_0$ and $\phi_1$ are the adjoint complex scalar fields in the two nodes of the quiver. Therefore, henceforth we will denote with $\mathcal{C}_{f,p}^+$ the integrated correlator obtained from \eqref{Z2intcorr} where the two Coulomb branch operators are untwisted and with $\mathcal{C}_{f,p}^-$ the integrated correlator with twisted chiral primary operators. By studying the integrated correlators $\mathcal{C}_{f,p}^+$ and $\mathcal{C}_{f,p}^-$ using localization, we are able to find an exact expression in the 't Hooft coupling in the leading planar approximation. From it we then obtain the leading strong-coupling behavior of these integrated correlators making use of techniques similar to
those developed in \cite{Belitsky:2020qir,Belitsky:2020qrm}. 

The second purpose of this paper is to examine the integrated correlators involving four moment map operators $\mathcal{J}(x)$, namely we study the quantities
\begin{subequations}
\begin{align}
& \mathcal{Y}_{f} \equiv \partial_{m_{f}}^4 \log\mathcal{Z}(m_1,m_2)\Big\vert_{m_1=m_2=0}  \, , \label{4mZ2} \\[0.5em]
& \mathcal{Y}_{1,2} \equiv \partial^2_{m_1}\partial^2_{m_2}\log\mathcal{Z}(m_1,m_2)\Big\vert_{m_1=m_2=0} \label{4mZ2bis}\,  .
\end{align}
\label{4mY}
\end{subequations}
Specifically the correlator defined in \eqref{4mZ2} corresponds to the integrated 4-point function of moment-map operators all belonging to the same hypermultiplet, whereas the second one, defined in \eqref{4mZ2bis}, is the integrated 4-point correlator involving two moment map operators from each of the two bifundamental hypermultiplets.

Also in this case, using the interacting matrix model of the $\mathbb{Z}_2$ quiver we manage to find expressions for these correlators which are valid for all values of the coupling in the leading and next-to-leading orders of the large-$N$ expansion and in both cases, building on the techniques of \cite{Belitsky:2020qir,Belitsky:2020qrm}, we idenitify the leading term of the strong-coupling expansion of these integrated four-point functions.

\subsection{Outline}

The rest of the paper is organised as follows. In Section \ref{sec:matrixmodel} we review the matrix model of the $\mathbb{Z}_2$ quiver gauge theory and find an exact expression for the first and the second mass corrections of the mass-deformed free energy which is valid for all values of the 't Hooft coupling in the planar and next-to-planar limits of the theory. In Section \ref{sec:2massder} we focus on the study of the mixed integrated correlators \eqref{Z2intcorr}. %involving two untwisted or twisted Coulomb branch operators of generic conformal dimension and two moment map operators. 
Using the matrix model representations of these correlators, we manage to derive their leading strong-coupling behavior in the large-$N$ limit. Then in Section \ref{sec:JJJJ} we examine the integrated correlators \eqref{4mY} in the planar and next-to-planar limits and %of four moment map operators. Exploiting the result obtained in Section \ref{sec:matrixmodel} for the second mass correction of the mass-deformed free energy of the $\mathbb{Z}_2$ quiver, we straightforwardly get the planar and next-to-planar expansions of this correlation function. Subsequently, after an intricate computation for each of the terms appearing in these expressions, we finally 
obtain their leading strong-coupling behavior. We draw our conclusions in Section \ref{conclusion} with an overview of our final results and a future outlook, mentioning open problems that could be further investigated starting from our analysis.

Moreover, many technical details about the matrix model computations and the analytical and numerical techniques, used to evaluate the strong-coupling behavior of the integrated correlators, are included in six appendices.

\section{The \texorpdfstring{$\mathbb{Z}_2$}{} quiver gauge theory and its massive deformation}
\label{sec:matrixmodel}
\begin{figure}[ht]
    \centering
\includegraphics[scale=1.0]{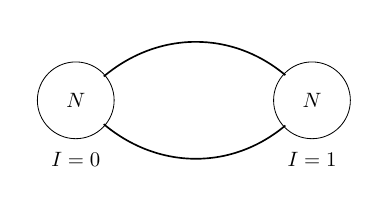}
\caption{The $4d$ $\mathcal{N}=2$ quiver gauge theory.}
    \label{fig:quiverZ2}
\end{figure}

We consider a massive deformation of the $4d$ $\mathcal{N}=2$ quiver gauge theory, which arises as a $\mathbb{Z}_2$ orbifold of $\mathcal{N}=4$ SYM with gauge group $SU(2N)$. The matter content of this theory can be represented by the quiver diagram reported in Figure \ref{fig:quiverZ2}. We label the nodes of the quiver with the index $I=0,1$. Each node corresponds to an $SU(N)$ gauge group and each line represents an $\mathcal{N}=2$ hypermultiplet transforming in the bifundamental representation.

This theory is conformal, since the  $\beta$-function vanishes for both nodes, and it possesses an $SU(2)_R \times U(1)_R$ R-symmetry group. In this work we solely focus on the most symmetric configuration, where the coupling constants associated with the two gauge nodes are equal, namely $g_0 = g_1 =g$. This leads us to organise the large-$N$ limit with a unique 't Hooft coupling $\lambda = g^2N$.
Using supersymmetric localization, the partition function $\mathcal{Z}(m_1,m_2)$ of the mass-deformed theory takes the following expression \cite{Billo:2021rdb}
\begin{align}
& \mathcal{Z}(m_1,m_2) = \left(\frac{8\pi^2N}{\lambda}\right)^{N^2-1}\int da\, \int\, db\, \, \text{e}^{-\frac{8\pi^2N}{\lambda}(\text{tr}\,a^2+\text{tr}\,b^2)}\,\Big|\mathcal{Z}_{\text{1-loop}}(a,b,m_1,m_2)\,\mathcal{Z}_{\text{inst}}(a,b,m_1,m_2,\lambda)\Big|^2 \, ,  
\label{Zm}
\end{align}
where we integrate over two Hermitian traceless matrices, $a$ and $b$, associated with the two nodes, employing the following notation
\begin{align}
a = a_c\,T^c\,, \ \ \ \ b=b_d\,T^d\,,   \ \ \ \  da = \prod_{c=1}^{N^2-1}\frac{da^c}{\sqrt{2\pi}}\,, \ \ \ \ db = \prod_{c=1}^{N^2-1}\frac{db^c}{\sqrt{2\pi}} 
\end{align}
with $c,d=1,\dots, N^2-1$ and $T_c$ are the $\mathfrak{su}(N)$  Lie algebra generators satisfying $\text{tr}\,T_cT_d = \frac{1}{2}\delta_{c,d}$. Moreover, $\mathcal{Z}_{\text{1-loop}}$ and $\mathcal{Z}_{\text{inst}}$ denote the perturbative 1-loop contribution and the instanton contribution, respectively. Henceforth, we solely focus on the large-$N$ limit of the theory at fixed $\lambda$, where the instanton contribution is exponentially suppressed; therefore we set $\mathcal{Z}_{\text{inst}}(a,b,m_1,m_2,\lambda) = 1$. On the other hand, the 1-loop contribution reads
\begin{equation}
\Big|\mathcal{Z}_{\text{1-loop}}(a,b,m_1,m_2)\Big|^2 = \frac{\displaystyle \prod_{k<\ell}^{N}H^{2}(a_k-a_\ell)\,H^2(b_k-b_\ell)}{ \displaystyle \prod_{f=1}^{2}\prod_{k,\ell=1}^{N}H(a_k-b_\ell+m_{f})^{1/2}\,H(a_k-b_\ell-m_{f})^{1/2}} \, ,  
\label{Z1loop}
\end{equation}
where the function $H(x)$ is defined as follows
\begin{equation}
H(x)=\,\rme^{-(1+\gamma)x^2}\,G(1+\ii\,x)\,G(1-\ii\,x)
\label{His}
\end{equation}
and where $G$ denotes the Barnes $G$-function, while $\gamma$ denotes the Euler-Mascheroni constant. To compute the integrated correlators of interest, it is sufficient to expand the expression \eqref{Z1loop} for small values of $m_1$ and $m_2$ up to the order $O(m^4)$. We obtain
\begin{align}
& \Big|\mathcal{Z}_{\text{1-loop}}(a,b,m_1,m_2)\Big|^2 = \nonumber \\
& \Big|\mathcal{Z}_{\text{1-loop}}(a,b,0,0)\Big|^2 \text{Exp}\left[1-\sum_{f=1}^{2} \frac{m_f^2}{2}\sum_{k,\ell=1}^{N}\partial^2\log\, H( a_k-  b_{\ell})  -\sum_{f=1}^{2}\frac{m_f^4}{24}\sum_{k,\ell=1}^{N} \partial^{4}\log H(a_{k}-b_{\ell}) + O(m^6) \right]\, .
\label{Z1loopExpansion}
\end{align}
The expression \eqref{Z1loopExpansion} can be regarded as an interaction action added to the Gaussian matrix model, consisting of an $m$-independent term and two massive contributions. In this manner, the small mass expansion of the massive partition function \eqref{Zm} can be written as
\begin{align}
\mathcal{Z}(m_1,m_2) = \left(\frac{8\pi^2N}{\lambda}\right)^{N^2-1}\int da\,\int\, db\, \, \text{e}^{-\frac{8\pi^2N}{\lambda}(\text{tr}\,a^2+\text{tr}\,b^2)}\,\text{e}^{-S_0-S_2-S_4 \, +\,  O(m^6)}\, ,  
\label{ZmExpansion}
\end{align}
where
\begin{subequations}
\begin{align}
& S_0 = -\sum_{k \neq \ell}^{N}\Big[\,\log H(a_k-a_\ell) + \log H(b_k-b_{\ell})\,\Big] +2\sum_{k,\ell=1}^{N}\log H(a_k-b_{\ell})\, , \label{S0}\\
& S_2 = \sum_{f=1}^{2}\frac{m_f^2}{2}\sum_{k,\ell=1}^{N}\partial^2 \log\, H(a_k-b_\ell)\, , \label{S2} \\
& S_4 = \sum_{f=1}^{2}\frac{m_f^4}{24}\sum_{k,\ell=1}^{N}\partial^4 \log H(a_k-b_\ell)\, .
\label{S4}
\end{align}
\label{Sall}
\end{subequations}
Then we can derive the small mass expansion of the mass-deformed free energy $\mathcal{F}(m_1,m_2)\equiv -\log\,\mathcal{Z}(m_1,m_2)$ up to the order $O(m^4)$, which reads
\begin{align}
\mathcal{F}(m_1,m_2) = \mathcal{F} + \, \langle S_2 \rangle  + \, \langle S_4 \rangle + \frac{1}{2}\left(\langle S_2 \rangle^2 - \langle S_2^2\rangle\right) \, + O(m^6) \, ,  
\label{FmExpansion}
\end{align}
where $\mathcal{F}\equiv\mathcal{F}(0,0)$ and the expectation value $\langle \cdot \rangle$ has been computed with respect to the massless matrix model. For a generic function $f$ this expectation value is defined as
\begin{align}
\langle f \rangle \, = \, \frac{ \displaystyle\int da \int db \, \text{e}^{-\frac{8\pi^2N}{\lambda}(\text{tr}\,a^2+\text{tr}\,b^2)}\,f\,\text{e}^{-S_0}}{ \displaystyle \int da \int db\, \text{e}^{-\frac{8\pi^2N}{\lambda}(\text{tr}\,a^2+\text{tr}\,b^2)}\,\text{e}^{-S_0}} \, = \, \frac{\langle f\, \text{e}^{-S_0}\ \rangle_0}{\langle \text{e}^{-S_0}\rangle_0} \ ,  
\end{align}
where $\langle \cdot \rangle_0$ denotes the vacuum expectation value in the Gaussian matrix model; these types of VEVs can be efficiently evaluated by exploiting the set of recursion relations satisfied by the following functions \cite{Billo:2017glv,Beccaria:2020hgy}
\begin{align}
t_{n_1,n_2,\dots,n_p} \equiv \langle \text{tr}\,a^{n_1}\,\text{tr}\,a^{n_2}\,\cdots\,\text{tr}a^{n_p}\rangle_0 \,  ,
\label{tt}
\end{align}
where $a$ denotes a matrix in the $\mathfrak{su}(N)$ Lie algebra.

To compute the quantities appearing on the right-hand side of \eqref{FmExpansion}, we begin by rescaling the matrices $a$ and $b$ as follows
\begin{align}
a \mapsto \sqrt{\frac{\lambda}{8\pi^2N}}\,a\,, \ \ \ \ \ \ \ \ b \mapsto \sqrt{\frac{\lambda} {8\pi^2N}}\,b \,  .
\end{align}
Then, as a second step, we use the small $x$ expansion of
\begin{align}
\log H(x) = \sum_{n=1}^{\infty}(-1)^n\,\frac{\zeta_{2n+1}}{n+1}x^{2n+2}\, ,   
\end{align}
where $\zeta_k$ denotes the Riemann zeta function $\zeta(k)$. This way, the expressions \eqref{Sall} are expressed as a power series in $\lambda$ and are given by 
\begin{gather}
 S_0 = \sum_{m=2}^{\infty}\sum_{k=2}^{2m}(-1)^{m+k}\left(\frac{\lambda}{8\pi^2N}\right)^{m}\left(\begin{array}{c}2m\\
 k\end{array}\right)\frac{\zeta_{2m-1}}{m}\left(\text{tr}\,a^{2m-k}-\text{tr}\,b^{2m-k}\right)\left(\text{tr}\,a^k-\text{tr}\,b^k\right)\,, \label{S0Expansion}\\[0.5em]
S_2 =-\sum_{f=1}^{2}m_f^2\,\mathcal{M}_{\mathbb{Z}_2}\,, \qquad \qquad  S_4 = -\sum_{f=1}^{2}m_f^4\,\mathcal{H}_{\mathbb{Z}_2}\,, 
\end{gather}
where
\begin{align}
& \mathcal{M}_{\mathbb{Z}_2} =-\sum_{n=1}^{\infty}\sum_{\ell=0}^{2n}(-1)^{n+\ell}\,\frac{(2n+1)!\,\zeta_{2n+1}}{(2n-\ell)!\,\ell!}\left(\frac{\lambda}{8\pi^2N}\right)^n\text{tr}\,a^\ell\,\text{tr}\,b^{2n-\ell} \, , \label{M}
\\[0.5em]
& \mathcal{H}_{\mathbb{Z}_2} =  - \frac{1}{24}\sum_{n=1}^{\infty}\sum_{\ell=0}^{2n-2}(-1)^{n+\ell}\,\frac{(2n+1)!\,\zeta_{2n+1}}{(2n-2-\ell)!\,\ell!}\left(\frac{\lambda}{8\pi^2N}\right)^{n-1}\text{tr}\,a^{\ell}\,\text{tr}\,b^{2n-2-\ell} \, \ .
\label{H}
\end{align}
Thus, upon examining the expressions \eqref{M} and \eqref{H}, it becomes evident that evaluating \eqref{FmExpansion} involves computing vacuum expectation values of the form $\langle \text{tr}\,a^{p}\,\text{tr}\,b^q \rangle$ in the (massless) interacting matrix model. In the next section, we will review how such computations can be efficiently carried out in the large-$N$ limit of the $\mathbb{Z}_2$ quiver gauge theory.

\subsection{The large-\texorpdfstring{$N$}{} expansion of the operators \texorpdfstring{$\mathcal{M}_{\mathbb{Z}_2}$}{} and \texorpdfstring{$\mathcal{H}_{\mathbb{Z}_2}$}{}}
As a first step, following \cite{Billo:2021rdb}, we introduce the untwisted (+) and  twisted (-) combinations 
\begin{align}
\label{AA}
A_{k}^{\pm} = \frac{1}{\sqrt{2}}(\text{tr}\,a^k \pm \text{tr}\,b^k)\, .  
\end{align}
Then, as shown in \cite{Billo:2022fnb}, it is advantageous to express these combinations in terms of a different operator basis, $\mathcal{P}_k^{\pm}$, defined by the following relation
\begin{align}
\label{PP}
A_{k}^{\pm} = \left(\frac{N}{2}\right)^{\frac{k}{2}}\sum_{\ell=0}^{\lfloor\frac{k-1}{2}\rfloor}\sqrt{k-2\ell}\left(\begin{array}{c}
     k  \\
     \ell 
\end{array}\right)\mathcal{P}_{k-2\ell}^{\pm}   + \langle A_{k}^{\pm} \rangle_0 \, .
\end{align}
We collect the main properties of the operators $\mathcal{P}_{k}^{\pm}$ in Appendix \ref{app:Pcorrelators}. Here, we just observe that in the free theory and in the large-$N$ limit, these operators are particularly useful since they are orthonormal, namely\footnote{We warn the reader that, from this point forward, the symbol $\simeq$ means that equality holds at the leading term of the large-$N$ expansion.}
\begin{align}
\langle \mathcal{P}_{k}^{\pm}\,\mathcal{P}_{\ell}^{\pm} \rangle_0 \, \simeq \, \delta_{k,\ell} \, .
\end{align}
Furthermore, the interaction action \eqref{S0Expansion} can be expressed only in terms of the twisted operators taking the form \cite{Billo:2021rdb,Billo:2022fnb}
\begin{align}
S_0 \,=\, -\frac{1}{2}\sum_{k,\ell=1}^{\infty}\mathcal{P}_{k}^{-}\,\textsf{X}_{k,\ell}\,\mathcal{P}_{\ell}^{-}\, \ ,
\label{S0viaX}
\end{align}
where the $\textsf{X}$-matrix is given by
\begin{align}
\textsf{X}_{k,\ell} \,=\, 2\,(-1)^{\frac{k+\ell+2k\ell}{2}+1}\sqrt{k\,\ell}\int_0^{\infty} \frac{dt}{t}\,\frac{1}{\sinh(t/2)^2}\,J_k\left(\frac{t\sqrt{\lambda}}{2\pi}\right)J_{\ell}\left(\frac{t\sqrt{\lambda}}{2\pi}\right)
\label{Xmatrix}
\end{align}
for $k,\ell \geq 2$ and its entries with opposite parities vanish, namely $\textsf{X}_{2k,2\ell+1} =0$. Importantly, we notice that all the dependence on the 't Hooft coupling, which in \eqref{S0Expansion} was expressed as a perturbative series, has been completely summed up in terms of Bessel functions of the first kind. For later convenience, we also introduce the matrices $\textsf{X}^{\text{odd}}$ and $\textsf{X}^{\text{even}}$, which are defined as
\begin{align}
(\textsf{X}^{\text{odd}})_{k,\ell} \,=\, \textsf{X}_{2k+1,2\ell+1}\,, \qquad (\textsf{X}^{\text{even}})_{k,\ell} \,=\, \textsf{X}_{2k,2\ell} \, .  \label{Xall}
\end{align}
As shown in \cite{Billo:2021rdb}, the relation \eqref{S0viaX} allows to express the partition function of the massless theory as
\begin{align}
\mathcal{Z} = [\text{det}(\mathbb{1}-\textsf{X})]^{-\frac{1}{2}}\, ,  
\end{align}
which, in turn, implies that the free energy can be expressed in terms of the \textsf{X}-matrix as
\begin{align}
\mathcal{F} = \frac{1}{2}\,\text{Tr}\,\log(\mathbb{1}-\mathsf{X})\, .    
\end{align}
In the following, we also need the expressions of  2-  and 3-point functions involving $\mathcal{P}_{k}^{\pm}$ operators in the interacting theory. In the large-$N$ limit, these correlators can be resummed in terms of the $\textsf{X}$-matrix as \cite{Billo:2022fnb}
\begin{align}
& \langle \mathcal{P}_{k}^{-}\,\mathcal{P}_{\ell}^{-} \rangle \simeq \textsf{D}_{k,\ell}\ , \\
&  \langle \mathcal{P}_{n_1}^{+}\,\mathcal{P}_{n_2}^{-}\,\mathcal{P}_{n_3}^{-} \rangle \, \simeq \, \frac{\sqrt{n_1}\,\textsf{d}_{n_2}\,\textsf{d}_{n_3}}{\sqrt{2}N} +\langle \mathcal{P}_{n_2}^{-}\,\mathcal{P}_{n_3}^{-} \rangle \, \langle \mathcal{P}_{n_1}^{+} \rangle \,  ,
\end{align}
where
\begin{align}
\textsf{D}_{k,\ell} = \delta_{k,\ell} + \textsf{X}_{k,\ell} + (\textsf{X}^2)_{k,\ell} + (\textsf{X}^3)_{k,\ell} + \dots \qquad \text{and} \qquad \textsf{d}_k = \sum_{\ell=2}^{\infty}\sqrt{\ell}\,\textsf{D}_{k,\ell} \,.
\label{dandD}
\end{align}

Using the relations \eqref{AA} and \eqref{PP}, the first mass correction \eqref{S2} to the free energy can be completely expressed in terms of the $\mathcal{P}_{k}^{\pm}$ operators. After a long computation, whose details are collected in Appendix \ref{app:MandH}, we obtain
\begin{align}
\mathcal{M}_{\mathbb{Z}_2} \,=\, \mathcal{M}_{\mathbb{Z}_2}^{(0)} + \mathcal{M}_{\mathbb{Z}_2}^{(1)}+
\mathcal{M}_{\mathbb{Z}_2}^{(2)}\, ,
\label{Mexpansion}
\end{align}
where the three terms on the r.h.s. contain zero, one and two $\mathcal{P}_{k}^{\pm}$ operators, respectively, and read
\begin{subequations}
\label{M2full}
\begin{align}
& \mathcal{M}_{\mathbb{Z}_2}^{(0)} = N^2\,\mathsf{M}_{0,0} + \mathsf{M}_{1,1} -\frac{1}{6}\sum_{k=1}^{\infty}\sqrt{2k+1}\mathsf{M}_{1,2k+1} + O\left(N^{-2}\right) \,, \\
& \mathcal{M}_{\mathbb{Z}_2}^{(1)} = \sqrt{2}N\sum_{k=1}^{\infty}\mathsf{M}_{0,2k}\mathcal{P}_{2k}^{+} + \frac{\sqrt{2}\,\lambda}{32\pi^2 N}\sum_{k=1}^{\infty}\mathsf{Q}_{0,2k}\mathcal{P}_{2k}^{+} - \frac{\lambda}{192\pi^2 N}\sum_{k=1}^{\infty}\mathsf{Q}_{2,2k}\,\mathcal{P}_{2k}^{+} +O(N^{-3})\,, \label{M21} \\
& \mathcal{M}_{\mathbb{Z}_2}^{(2)} = \frac{1}{2}\sum_{p,q=2}^{\infty}(-1)^{p-pq}\,\mathsf{M}_{p,q}\,(\mathcal{P}_{p}^{+}\,\mathcal{P}_{q}^{+}-\mathcal{P}_{p}^{-}\,\mathcal{P}_{q}^{-})\, ,   \end{align}
\end{subequations}
where the coefficients $\mathsf{M}_{n,m}$ are defined as follows
\begin{subequations}
\begin{align}
\mathsf{M}_{0,0}&=\int_0^\infty\!\frac{dt}{t}\,\frac{(t/2)^2}{\sinh(t/2)^2} \bigg[1-\frac{16\,\pi^2}{t^2\lambda}\,J_1\Big(\frac{t\sqrt{\lambda}}{2\pi}\Big)^2\bigg]~,\label{M00}\\[2mm]
\mathsf{M}_{0,n}&=(-1)^{\frac{n}{2}+1}\,\sqrt{n}\!\int_0^\infty\!\frac{dt}{t}\,\frac{(t/2)^2}{\sinh(t/2)^2}\, \Big(\frac{4\pi}{t\sqrt{\lambda}}\Big)\,J_1\Big(\frac{t\sqrt{\lambda}}{2\pi}\Big)\,J_n\Big(\frac{t\sqrt{\lambda}}{2\pi}\Big)~,\label{M0n}\\[2mm]
\mathsf{M}_{n,m}&=(-1)^{\frac{n+m+2nm}{2}+1}\,\sqrt{nm}\!\int_0^\infty\!\frac{dt}{t}\,\frac{(t/2)^2}{\sinh(t/2)^2}
\,J_n\Big(\frac{t\sqrt{\lambda}}{2\pi}\Big)\,J_m\Big(\frac{t\sqrt{\lambda}}{2\pi}\Big)\label{Mnm}\,  .
\end{align}
\label{Mmatrix}%
\end{subequations}
In analogy with \eqref{Xall}, we further introduce the matrices $\textsf{M}^{\text{odd}}$ and $\textsf{M}^{\text{even}}$, originating from \eqref{Mnm}, and defined as
\begin{align}
(\textsf{M}^{\text{odd}})_{k,\ell} \,=\, \textsf{M}_{2k+1,2\ell+1}\,, \qquad (\textsf{M}^{\text{even}})_{k,\ell} \,=\, \textsf{M}_{2k,2\ell} \,  .  \label{Mall}
\end{align}
Finally, the coefficients $\textsf{Q}_{0,n}$ and $\textsf{Q}_{n,m}$, which appear in \eqref{M21} in the next-to-planar term of $\mathcal{M}_{\mathbb{Z}_2}^{(1)}$, read
\begin{subequations}
\begin{align}
& \textsf{Q}_{0,n} = (-1)^{\frac{n}{2}+1}\sqrt{n}\int_0^{\infty}\,dt\, t\, \frac{(t/2)^2}{\sinh(t/2)^2}\left(\frac{4\pi}{t\sqrt{\lambda}}\right)J_1\left(\frac{t\sqrt{\lambda}}{2\pi}\right)J_n\left(\frac{t\sqrt{\lambda}}{2\pi}\right)\, , \label{Q0n} \\[0.75em]
& \textsf{Q}_{n,m} = (-1)^{\frac{n+m+2nm}{2}+1}\sqrt{n\,m}\int_{0}^{\infty}\, dt\, t\, \frac{(t/2)^2}{\sinh(t/2)^2}J_n\left(\frac{t\sqrt{\lambda}}{2\pi}\right)J_m\left(\frac{t\sqrt{\lambda}}{2\pi}\right)\,  . \label{Qnm}
\end{align}
\label{Qmatrix}
\end{subequations}
In this case as well, for later convenience, we find it useful to introduce the matrices $\textsf{Q}^{\text{odd}}$ and $\textsf{Q}^{\text{even}}$, defined as
\begin{align}
(\textsf{Q}^{\text{odd}})_{k,\ell} = \textsf{Q}_{2k+1,2\ell+1}\, ,   \qquad (\textsf{Q}^{\text{even}})_{k,\ell} = \textsf{Q}_{2k,2\ell}\, .
\label{Qall}
\end{align}

Using the same procedure, we can further demonstrate that the operator $\mathcal{H}_{\mathbb{Z}_2}$ \eqref{H}, necessary for evaluating the second mass correction to the free energy \eqref{FmExpansion}, can also be expressed in terms of the $\mathcal{P}_{k}^{\pm}$ operators. This time, we obtain
\begin{align}
\mathcal{H}_{\mathbb{Z}_2} \,=\, \mathcal{H}_{\mathbb{Z}_2}^{(0)} + \mathcal{H}_{\mathbb{Z}_2}^{(1)}+ \mathcal{H}_{\mathbb{Z}_2}^{(2)} \, ,
\end{align}
where the three terms on the r.h.s. contain  zero, one and two $\mathcal{P}_{k}^{\pm}$ operators, respectively, and are given by \footnote{We refer to Appendix \ref{app:MandH} for the details of this computation.} 
\begin{subequations}
\label{H2full}
\begin{align}
& \mathcal{H}_{\mathbb{Z}_2}^{(0)} = -\frac{4\pi^2}{3\lambda}\textsf{M}_{1,1}N^2 - \frac{1}{12}\left(\textsf{Q}_{1,1}-\frac{1}{6}\sum_{k=1}^{\infty}\sqrt{2k+1}\textsf{Q}_{1,2k+1}\right) + O(N^{-2}) \, , \label{H0} \\
& \mathcal{H}_{\mathbb{Z}_2}^{(1)} = -\frac{\sqrt{2}N}{12}\sum_{k=1}^{\infty}\textsf{Q}_{0,2k}\,\mathcal{P}_{2k}^{+} + O(N^{-1})\,  , \label{H1} \\      
& \mathcal{H}_{\mathbb{Z}_2}^{(2)} = -\frac{1}{24}\sum_{p,q=2}^{\infty}(-1)^{p-pq}\,\textsf{Q}_{p,q}\,(\mathcal{P}^{+}_{p}\mathcal{P}^{+}_{q}-\mathcal{P}_{p}^{-}\mathcal{P}_{q}^{-}) \, . \label{H2}
\end{align}
\end{subequations}

\section{The two mass derivatives integrated correlators}
\label{sec:2massder}
In this section we perform the computation of the leading order of the large-$N$ expansion of the integrated correlators $\mathcal{C}_{f,p}^+$ and $\mathcal{C}_{f,p}^-$ defined in the Introduction. Following the same reasoning as \cite{Billo:2023kak}, in the 
$\mathbb{Z}_2$ quiver matrix model these two integrated correlators are represented respectively by
\begin{align}
& \mathcal{C}_{f,p}^{+} \,= \frac{1}{2}\left(\langle O_p^{+}\,O_p^{+}\,\mathcal{M}_{\mathbb{Z}_2} \rangle - \langle O^{+}_p\,O^{+}_p \rangle\,\langle \mathcal{M}_{\mathbb{Z}_2} \rangle\right)\, , \label{Cplus} \\[0.5em]
& \mathcal{C}_{f,p}^{-} \, = \frac{1}{2}\left(\langle O_p^{-}\,O_p^{-}\,\mathcal{M}_{\mathbb{Z}_2} \rangle - \langle O_p^{-}\,O_p^{-} \rangle\,\langle \mathcal{M}_{\mathbb{Z}_2} \rangle\right)\ ,
\label{Cminus}
\end{align}
where $O_p^+$ and $O_p^-$ are the quantities defined in the matrix model \cite{Billo:2021rdb,Billo:2022fnb} that corespond to the untwisted and twisted operators defined in \eqref{utCPO}, while the large-$N$ expansion of $\mathcal{M}_{\mathbb{Z}_2}$ is reported in \eqref{M2full}.
\subsection{The untwisted correlator} %$\mathcal{C}_p^{+}$}
As discussed in \cite{Billo:2022fnb}, the untwisted operators $\mathcal{O}_p^{+}$ in the matrix model are represented by
\begin{align}
O_p^{+} = \sqrt{\mathcal{G}_p^{(0)}}\left(\mathcal{P}^{+}_{p}-\langle \mathcal{P}^{+}_{p} \rangle\right)\, .  
\end{align}
where $\mathcal{G}_p^{(0)}$ denotes to the 2-point function in the free theory, namely\footnote{It is worth remembering that, for the untwisted operators $O_p^{+}$, as shown in \cite{Billo:2021rdb,Billo:2022fnb}, $\mathcal{G}_p^{(0)}$ corresponds also to the 2-point function in the interacting theory.}
\begin{align}
\langle O_p^{+}O_p^{+} \rangle_0 \, \simeq \, \mathcal{G}_p^{(0)} = p\left(\frac{N}{2} \right)^p \, .    
\end{align}

Thus, using the large-$N$ expansion \eqref{M2full} of the $\mathcal{M}_{\mathbb{Z}_2}$ operator, along with the results collected in Appendix \ref{app:Pcorrelators}, we compute the leading order of the large-$N$ expansion for $\mathcal{C}_{f,p}^{+}$ and we obtain
\begin{align}
& \mathcal{C}_{f,p}^{+}= \frac{1}{2}\left(\langle O^{+}_p\,O^{+}_p\,\mathcal{M}_{\mathbb{Z}_2} \rangle - \langle O^{+}_p\,O^{+}_p\, \rangle\,\langle \mathcal{M}_{\mathbb{Z}_2} \rangle  \right) \simeq  \frac{N}{\sqrt{2}}\sum_{k=0}^{\infty}\textsf{M}_{0,2k}\,\left(\langle O^{+}_p\,O^{+}_p\,\mathcal{P}_{2k}^{+} \rangle -\langle O^{+}_p\,O^{+}_p \rangle\langle \mathcal{P}_{2k}^+ \rangle\right) + \nonumber \\ 
&  \frac{1}{4}\sum_{k,\ell=2}^{\infty}(-1)^{k-k\,\ell}\,\textsf{M}_{k,\ell}\left[\langle O^{+}_p\,O^{+}_p\,\mathcal{P}^{+}_{k}\mathcal{P}^{+}_{\ell} \rangle - \langle O^{+}_p\,O^{+}_p\rangle\langle\mathcal{P}^{+}_{k}\,\mathcal{P}^{+}_{\ell} \rangle - \langle O^{+}_p\,O^{+}_p\,\mathcal{P}^{-}_{k}\,\mathcal{P}^{-}_{\ell} \rangle  + \langle O^{+}_p\,O^{+}_p\rangle\langle\mathcal{P}^{-}_{k}\,\mathcal{P}^{-}_{\ell}\rangle \right] \nonumber  \\
& = \frac{\mathcal{G}_p^{(0)}}{2}\left(p\sum_{k=0}^{\infty}\sqrt{2k}\,\mathsf{M}_{0,2k} + \,\mathsf{M}_{p,p} \right) = \frac{\mathcal{G}_p^{(0)}}{2}\left(\,\mathsf{M}_{p,p}-p\,\mathsf{M}_{1,1}\right)\, .
\end{align}
where in the last step we exploited the identity
\footnote{We refer to Appendix B of \cite{Billo:2023kak} for its proof.} 
\begin{align}
\sum_{k=1}^{\infty}\sqrt{2k}\,\textsf{M}_{0,2k} = -\textsf{M}_{1,1}\, .
\label{idM11}
\end{align}
Therefore, we finally find
\begin{align}
\frac{\mathcal{C}_{f,p}^{+}}{\mathcal{G}_p^{(0)}} = \frac{1}{2}\left(  \mathsf{M}_{p,p}-p\,\mathsf{M}_{1,1}\right) \, \ . 
\end{align}
We observe that this expression is valid for any value of the 't Hooft coupling and coincides with the $\mathcal{N}=4$ SYM result of \cite{Binder:2019jwn} up to a numerical multiplicative factor. In particular, we can evaluate it at strong-coupling by  applying the Mellin-Barnes method to the matrix \textsf{M} \eqref{Mnm}. Indeed, by writing the products between Bessel functions appearing in its definition as an
inverse Mellin transform
\begin{align}
J_n\left(\frac{t\sqrt{\lambda}}{2\pi}\right)\,J_m\left(\frac{\sqrt{\lambda}}{2\pi}\right) = \int_{-\text{i}\infty}^{+\text{i}\infty} \frac{ds}{2\pi\text{i}}\,\frac{\Gamma(-s)\Gamma(2s+n+m+1)}{\Gamma(s+n+1)\Gamma(s+m+1)\Gamma(s+n+m+1)}\,\left( \frac{t\sqrt{\lambda}}{4\pi} \right)^{2s+n+m}
\end{align}
and performing the integral over $t$ in \eqref{Mnm} making use of the identity
\begin{align}
\label{integralidentity}
(2n+1)!\,\zeta_{2n+1} = \int_0^\infty \frac{dt}{t}\,\frac{(t/2)^2}{\sinh(t/2)^2}t^{2n}  \,,
\end{align}
$\textsf{M}_{n,m}$ ends up to be written as an expression such that, as already shown \cite{Billo:2023kak}, its asymptotic expansion for large values of $\lambda$ receives contributions from the poles on the negative real axis and, by evaluating the residues over these poles, it is straightforward to find 
\begin{align}
\label{Mstrongcoupling}
\textsf{M}_{n,m} \underset{\lambda \rightarrow \infty}{\sim} -\frac{1}{2}\delta_{n,m} + \frac{\sqrt{n\,m}}{\sqrt{\lambda}} + O\left(\lambda^{-\frac{3}{2}}\right) \,.
\end{align}
From this result it immediately follows
\begin{align}
\label{finaluntwisted}
\frac{\mathcal{C}_{f,p}^{+}}{\mathcal{G}_p^{0}} \, \underset{\lambda \rightarrow \infty}{\sim} \, \frac{p-1}{4} \, +O\left(\lambda^{-\frac{1}{2}}\right) \, .
\end{align}

\subsection{The twisted correlator} %$\mathcal{C}_p^{-}$}
We begin by recalling that, as shown in \cite{Billo:2022fnb}, the twisted operators $\mathcal{O}_p^{-}$ in the matrix model are represented by
\begin{align}
O_p^{-} = 
\sqrt{\mathcal{G}_p^{(0)}}\left(\mathcal{P}^{-}_{p}-\sum_{q < p} \textsf{W}_{p,q}\,\mathcal{P}_q^-\right)\, ,
\label{twistedmatrixmodel}
\end{align}
where the coefficients $\textsf{W}_{p,q}$ depend on $\lambda$, and they can be explicitly computed by imposing orthogonality of the operator $O_p^{-}$ to all lower-dimensional twisted operators \cite{Billo:2022fnb}. Then, we evaluate the leading order of the large-$N$ expansion of the integrated correlator \eqref{Cminus} and we obtain
\begin{align}
\label{C-exact}
\mathcal{C}_{f,p}^{-} \, = \, \frac{\mathcal{G}_p^{(0)}}{2}\biggl( \Pi_{p,p} -2 \sum_{p'< p} \textsf{W}_{p,p'}\, \Pi_{p',p} +\sum_{p',p''< p} \textsf{W}_{p,p'}\, \Pi_{p',p''}\, \textsf{W}_{p,p''} \biggr) \,,
\end{align} 
where
\begin{align}
& \Pi_{p,q} \, \equiv \, \langle \mathcal{P}^-_p \mathcal{P}^-_q \mathcal{M}_{\mathbb{Z}_2} \rangle - \langle \mathcal{P}^-_p \mathcal{P}^-_q \rangle \langle \mathcal{M}_{\mathbb{Z}_2} \rangle \, \simeq \, \sqrt{2}N\sum_{k=1}^{\infty}\textsf{M}_{0,2k}\,\left(\langle \mathcal{P}^-_p \mathcal{P}^-_q \mathcal{P}_{2k}^{+} \rangle -\langle \mathcal{P}^-_p \mathcal{P}^-_q \rangle\,\langle \mathcal{P}_{2k}^+ \rangle\right) + \nonumber \\ 
&  \frac{1}{2}\sum_{k,\ell=2}^{\infty}(-1)^{k-k\ell}\,\textsf{M}_{k,\ell}\left[\langle \mathcal{P}^-_p\,\mathcal{P}^-_q\,\mathcal{P}^{+}_{k}\mathcal{P}^{+}_{\ell} \rangle - \langle \mathcal{P}^-_p\,\mathcal{P}^-_q\rangle\langle\mathcal{P}^{+}_{k}\,\mathcal{P}^{+}_{\ell} \rangle - \langle \mathcal{P}^-_p\,\mathcal{P}^-_q\,\mathcal{P}^{-}_{k}\,\mathcal{P}^{-}_{\ell} \rangle  + \langle \mathcal{P}^-_p\,\mathcal{P}^-_q\rangle\langle\mathcal{P}^{-}_{k}\,\mathcal{P}^{-}_{\ell}\rangle \right]\, .
\end{align}
We exploit the large-$N$ expressions for the correlation functions between $\mathcal{P}_k^\pm$ operators collected in  Appendix \ref{app:Pcorrelators} and, in this way, we find
\begin{align}
\label{Pi}
& \Pi_{p,q} \, \simeq \,  \, \mathsf{d}_p\,\mathsf{d}_q \sum_{k=1}^{\infty}\textsf{M}_{0,2k}\,\sqrt{2k} -\frac{1}{2} \sum_{k,\ell=2}^{\infty}(-1)^{k-k\ell}\,(\textsf{D}_{p,k}\,\textsf{M}_{k,\ell}\,\textsf{D}_{\ell,q} + \textsf{D}_{q,k}\,\textsf{M}_{k,\ell}\,\textsf{D}_{\ell,p} ) = \nonumber \\
& -\mathsf{d}_p\,\mathsf{d}_q \,\mathsf{M}_{1,1} -\sum_{k,\ell=2}^{\infty}\,\textsf{D}_{p,k}\,\textsf{M}_{k,\ell}\,\textsf{D}_{\ell,q}\,,
\end{align}
where in the last step we again used the \eqref{idM11}.
At strong-coupling, the expression \eqref{twistedmatrixmodel} simplifies and   reduces to \cite{Billo:2022fnb} 
\begin{align}
O^{-}_p \,  \underset{\lambda \rightarrow \infty}{\sim} \,\sqrt{\mathcal{G}_p^{(0)}}\,\biggl(\mathcal{P}^{-}_{p} - \sqrt{\frac{p}{p-2}}\,\mathcal{P}_{p-2}^-\biggr)\,,
\end{align}
therefore from \eqref{C-exact} it follows that
\begin{align}
\label{RatioCpminus}
\frac{\mathcal{C}_{f,p}^{-}}{\mathcal{G}_p^{(0)}} \, \underset{\lambda \rightarrow \infty}{\sim} \, \frac{1}{2}\,\biggl( \Pi_{p,p} -2\, \sqrt{\frac{p}{p-2}}\, \Pi_{p,p-2} + \frac{p}{p-2}\,\Pi_{p-2,p-2}  \biggr)\,.
\end{align}
Moreover it holds that \cite{Billo:2022fnb}
\begin{align}
\mathsf{d}_k \, \underset{\lambda \rightarrow \infty}{\sim} \, \frac{\pi}{\sqrt{\lambda}} \biggl[ \frac{\sqrt{k}}{2}\bigl( k^2-\delta_{k\,\text{mod}\,2,1} \bigr) \biggr]  + O(\lambda^{-1}) \,,
\end{align}
hence we get
\begin{align}
-\mathsf{d}_p\,\mathsf{d}_q \,\mathsf{M}_{1,1} \,\underset{\lambda \rightarrow \infty}{\sim} \, \frac{\pi^2}{8\lambda}\,\sqrt{p\,q}\,(p^2-\delta_{p\,\text{mod}\,2,1})\,(q^2-\delta_{q\,\text{mod}\,2,1}) + O(\lambda^{-\frac{3}{2}})\,  ,
\end{align}
while, as argued in \cite{Billo:2023kak}, the other term in \eqref{Pi} scales as
\begin{align}
(\textsf{D}\,\textsf{M}\,\textsf{D})_{p,q} \underset{\lambda \rightarrow \infty}{\sim} O(\lambda^{-3/2}) \, ,
\end{align}
therefore we can neglect this last contribution and we approximate the expression \eqref{Pi} as
\begin{align}
\Pi_{p,q} \simeq -\,\textsf{d}_{p}\,\textsf{d}_{q}\,\textsf{M}_{1,1} \, .
\end{align}
Then, the ratio \eqref{RatioCpminus} becomes
\begin{align}
\frac{\mathcal{C}_{f,p}^{-}}{\mathcal{G}_{p}^{(0)}} \, \simeq \, -\frac{\textsf{M}_{1,1}}{2}\left(\textsf{d}_p - \sqrt{\frac{p}{p-2}}\textsf{d}_{p-2}\right)^2 \, \underset{\lambda \rightarrow \infty}{\sim} \, \frac{1}{4}\,\left(\frac{2\pi}{\sqrt{\lambda}}\,\sqrt{p}\,(p-1)\right)^2 \,  .
\end{align}
We recall that the ratio between the 2-point function $\mathcal{G}_p$ of twisted chiral primary operators in the $\mathbb{Z}_2$ quiver gauge theory and their counterpart in $\mathcal{N}=4$ SYM reads \cite{Billo:2022fnb}
\begin{align}
\frac{\mathcal{G}_p}{\mathcal{G}_p^{(0)}} \underset{\lambda \rightarrow \infty}{\sim} \frac{4\pi^2p\,(p-1)}{\lambda} \, +\, O(\lambda^{-3/2})\, . 
\end{align}
Exploiting this last relation we finally find
\begin{align}
\frac{\mathcal{C}_{f,p}^{-}}{\mathcal{G}_p} \,\underset{\lambda \rightarrow \infty}{\sim} \, \frac{p-1}{4} \, +\,  O(\lambda^{-1/2})\, .    
\end{align}
This is formally analogous to the result obtained for the untwisted integrated correlator in \eqref{finaluntwisted}. Furthermore, it is what occurs also in another example of $\mathcal{N}=2$ SCFT, namely the \textbf{E}-theory \cite{Billo:2023kak}, as well as in $\mathcal{N}=4$ SYM \cite{Binder:2019jwn}.

\section{The four mass derivatives integrated correlator 
}
\label{sec:JJJJ}
In this section we evaluate the leading and next-to-leading terms of the large-$N$ expansion of the integrated correlators defined in \eqref{4mY}. We begin by noting that, once it is expressed in terms of the expectation values of the operators $\mathcal{H}_{\mathbb{Z}_2}$ and $\mathcal{M}_{\mathbb{Z}_2}$, the second mass correction to the free energy \eqref{FmExpansion} is given by
\begin{align}
-\sum_{f=1}^{2}m_f^4\,\langle \mathcal{H}_{\mathbb{Z}_2} \rangle-\frac{1}{2}\left(\sum_{f=1}^{2}m_f^2\right)^2\left(\langle \mathcal{M}^2_{\mathbb{Z}_2} \rangle-\langle \mathcal{M}_{\mathbb{Z}_2} \rangle^2\right)\, \ .
\label{mToThe4}
\end{align}

%\begin{align}
% - m^4\left[\langle \mathcal{
% H}_{\mathbb{Z}_2} \rangle +\frac{1}{2}\left(\langle \mathcal{M}_{\mathbb{Z_2}}^2 \rangle -\langle \mathcal{M}_{\mathbb{Z}_2} \rangle^2 \right) \right] \, .
%\label{mToThe4}
%\end{align}
Thus, inserting this result in the definition \eqref{4mY}, the integrated correlators become
\begin{subequations}
\begin{align}
& \mathcal{Y}_f = 24 \left[\langle \mathcal{
 H}_{\mathbb{Z}_2} \rangle +\frac{1}{2}\left(\langle \mathcal{M}_{\mathbb{Z_2}}^2 \rangle -\langle \mathcal{M}_{\mathbb{Z}_2} \rangle^2 \right) \right] \, , \\
& \mathcal{Y}_{1,2} = 4\left(\langle \mathcal{M}_{\mathbb{Z_2}}^2 \rangle -\langle \mathcal{M}_{\mathbb{Z}_2} \rangle^2 \right)\, \ . 
\end{align}
\label{JJJJ}
\end{subequations}
It is easy to see that the expectation value of the operator $\mathcal{H}_{\mathbb{Z}_2}$ at the leading and next-to-leading orders of the large-$N$ expansion can be computed using the \eqref{H2full} and the expressions collected in Appendix \ref{app:Pcorrelators} and it reads
\begin{align}
 \langle \mathcal{H}_{\mathbb{Z}_2} \rangle = & -\frac{4\pi^2}{3\lambda}\textsf{M}_{1,1}N^2 - \frac{1}{12}\left(\textsf{Q}_{1,1}-\frac{1}{6}\sum_{k=1}^{\infty}\sqrt{2k+1}\textsf{Q}_{1,2k+1}\right) -\frac{N}{6\sqrt{2}}\sum_{k=1}^{\infty}\textsf{Q}_{0,2k}\,\langle \mathcal{P}_{2k}^{+} \rangle \nonumber \\
& -\frac{1}{24}\sum_{p,q=2}^{\infty}(-1)^{p-pq}\,\textsf{Q}_{p,q}\,(\langle \mathcal{P}^{+}_{p}\mathcal{P}^{+}_{q}\rangle-\langle \mathcal{P}_{p}^{-}\mathcal{P}_{q}^{-}\rangle)\, . \label{HvevFull}   
\end{align}
On the other hand, it is useful to express the difference between  $\langle \mathcal{M}_{\mathbb{Z_2}}^2 \rangle $ and $\langle \mathcal{M}_{\mathbb{Z_2}} \rangle^2$ appearing in \eqref{mToThe4} in terms of the connected correlators
\begin{align}
\langle \mathcal{M}_{\mathbb{Z}_2}\,\mathcal{M}^{'}_{\mathbb{Z}_2} \rangle_{\text{con}} \equiv \langle \mathcal{M}_{\mathbb{Z}_2}\,\mathcal{M}^{'}_{\mathbb{Z}_2} \rangle - \langle \mathcal{M}_{\mathbb{Z}_2} \rangle \, \langle \mathcal{M}^{'}_{\mathbb{Z}_2} \rangle \, .
\label{Mconnected}
\end{align}
Therefore, using the expansion \eqref{Mexpansion}, we obtain
\begin{align}
& \langle \mathcal{M}_{\mathbb{Z}_2}^2 \rangle -\langle \mathcal{M}_{\mathbb{Z}_2} \rangle^2   = \sum_{j=0}^{2}\langle \mathcal{M}^{(j)\,2}_{\mathbb{Z}_2} \rangle_{\text{con}}  +2\left(\langle \mathcal{M}^{(0)}_{\mathbb{Z}_2}\,\mathcal{M}^{(1)}_{\mathbb{Z}_2} \rangle_{\text{con}} +\langle \mathcal{M}^{(0)}_{\mathbb{Z}_2}\,\mathcal{M}^{(2)}_{\mathbb{Z}_2} \rangle_{\text{con}} +\langle \mathcal{M}^{(1)}_{\mathbb{Z}_2}\,\mathcal{M}^{(2)}_{\mathbb{Z}_2} \rangle_{\text{con}}\right) \,  .
\label{M2difference}
\end{align}
Then, using the expressions \eqref{M2full}, it is easy to check that 
\begin{align}
\langle \mathcal{M}^{(0)\,2}_{\mathbb{Z}_2} \rangle_{\text{con}} = \langle \mathcal{M}^{(0)}_{\mathbb{Z}_2}\,\mathcal{M}^{(1)}_{\mathbb{Z}_2} \rangle_{\text{con}} = \langle \mathcal{M}^{(0)}_{\mathbb{Z}_2}\,\mathcal{M}^{(2)}_{\mathbb{Z}_2} \rangle_{\text{con}} = 0\,  ,
\end{align}
while the three non-trivial contributions present in \eqref{M2difference} are given by
\begin{subequations}
\begin{align}
& \langle \mathcal{M}^{(1)\,2}_{\mathbb{Z}_2} \rangle_{\text{con}} \, \simeq \, 2 N^2\sum_{k=1}^{\infty}(\textsf{M}_{0,2k})^2 + 2\sum_{k=1}^{\infty}\sum_{\ell=1}^{\infty}\textsf{M}_{0,2k}\textsf{M}_{0,2\ell}\textsf{T}^{(+)}_{k,\ell}  \nonumber\\
&  \qquad\qquad\qquad +\frac{\lambda}{8\pi^2}\sum_{k=1}^{\infty}\textsf{M}_{0,2k}\textsf{Q}_{0,2k} - \frac{\sqrt{2}\,\lambda}{96\pi^2}\sum_{k=1}^{\infty}\textsf{M}_{0,2k}\textsf{Q}_{2,2k} -\,\textsf{M}_{1,1}^2(\lambda\partial_{\lambda}\mathcal{F})^2 + O(N^{-2})\,  , \label{M1connected} \\[0.5em] 
& \langle \mathcal{M}^{(2)\,2}_{\mathbb{Z}_2} \rangle_{\text{con}} \, \simeq  \frac{1}{2}\,\left(\text{Tr}[(\textsf{M}^{\text{even}})^2]+\text{Tr}[(\textsf{M}^{\text{odd}})^2] \right.\nonumber \\
&\left. \qquad\qquad\qquad +\text{Tr}[\textsf{M}^{\text{even}}\,\textsf{D}^{\text{even}}\,\textsf{M}^{\text{even}}\,\textsf{D}^{\text{even}}]+\text{Tr}[\textsf{M}^{\text{odd}}\,\textsf{D}^{\text{odd}}\,\textsf{M}^{\text{odd}}\,\textsf{D}^{\text{odd}}]\right) +O(N^{-2})\, , \label{M2connected} \\[0.5em]
& 2\langle \mathcal{M}^{(1)}_{\mathbb{Z}_2}\mathcal{M}^{(2)}_{\mathbb{Z}_2} \rangle_{\text{con}} \, \simeq \,  - \textsf{M}_{1,1}\sum_{p,q=2}^{\infty}(-1)^{p-pq}\,\textsf{M}_{p,q}(\sqrt{pq}-\textsf{d}_p\textsf{d}_q) \nonumber \\
& \qquad\qquad\qquad \qquad \ \ -2(\lambda\partial_{\lambda}\mathcal{F})  \sum_{k=1}^{\infty}\sum_{p=1}^{\infty}\textsf{M}_{0,2k}\textsf{M}_{2k,2p}\sqrt{2p} +O(N^{-2})\, , 
\label{M1M2con}
\end{align}
\label{MConnectedFull}
\end{subequations}
where we used the results of Appendix \ref{app:Pcorrelators} and the identity \eqref{idM11}, while the symbol $T^{+}_{k,\ell}$ appearing in the first line of \eqref{M1connected} denotes the next-to-leading order of the large-$N$ expansion of the correlator $\langle \mathcal{P}^{+}_{2k}\,\mathcal{P}^{+}_{2\ell}  \rangle$, namely
\begin{align}
\langle \mathcal{P}^{+}_{2k}\,\mathcal{P}^{+}_{2\ell}  \rangle \, = \, \delta_{k,\ell} \, +\,  \frac{\textsf{T}^{+}_{k,\ell}}{N^2} \, + \, O(N^{-4}) \,  .    \label{PPplus}
\end{align}
We argue in Appendix \ref{app:Tplus} that such  term is given by
\begin{align}
\textsf{T}_{k,\ell}^{+} = \sqrt{k\,\ell}\left(\frac{(k^2+\ell^2-1)(k^2+\ell^2-14)}{12}-(k^2+\ell^2-1)\lambda\partial_{\lambda}\mathcal{F} - (\lambda\partial_\lambda)^2\mathcal{F} + (\lambda\partial_{\lambda}\mathcal{F})^2 \right)\,  .
\label{Tplus}
\end{align}
It is worth noting that all the dependence on the 't Hooft coupling appears solely through a differential operator acting on the free energy of the massless theory.

By substituting the expressions \eqref{HvevFull} and \eqref{MConnectedFull} into  \eqref{JJJJ}, we derive the exact expression for the integrated correlators \eqref{JJJJ} in terms of the 't Hooft coupling. In the next two sections, we analyse separately the planar and next-to-planar contributions and their large-$\lambda$ expansions.

\subsection{The planar term}
Upon a brief examination of the expressions \eqref{HvevFull} and \eqref{MConnectedFull}, it becomes evident that the leading term of the large-$N$ expansion of the integrated correlators \eqref{JJJJ} read
\begin{subequations}
\begin{align}
& \mathcal{Y}_f \, \simeq \, -24N^2\left[\frac{4\pi^2}{3\lambda}\textsf{M}_{1,1}-\sum_{k=1}^{\infty}(\textsf{M}_{0,2k})^2\right]\, , \label{YfLO} \\
& \mathcal{Y}_{1,2} \, \simeq \, 8N^2\sum_{k=1}^{\infty}(\textsf{M}_{0,2k})^2\, \ .
\end{align}
\label{JJJJLO}
\end{subequations}
Then, using the explicit expressions \eqref{M2full} for the $\textsf{M}_{1,1}$ and $\textsf{M}_{0,2k}$ matrices as convolutions between Bessel functions of the first kind, it can be readily demonstrated that the above expression \eqref{YfLO} differs from the planar limit of its $\mathcal{N}=4$ SYM counterpart  $\mathcal{Y}^{(0)}$ computed in \cite{Chester:2020dja,Alday:2023pet}.

For completeness, it is worth noting that the strong-coupling expansion of the first term in \eqref{YfLO} comes out straightforwardly from \eqref{Mstrongcoupling}, whereas the second term in \eqref{YfLO} can be  obtained by employing the method introduced in Appendix A of \cite{Alday:2023pet}. Therefore, the first few terms of the strong coupling expansions of the correlators \eqref{JJJJLO} are

\begin{subequations}
\begin{align}
\mathcal{Y}_f \, & \underset{\lambda \rightarrow \infty}{\sim} \,
N^2\left[3+\frac{8\pi^2}{\lambda}+\frac{24}{\lambda^{3/2}}\left(2\zeta_3-\frac{2 \pi ^2}{3}\right)+\frac{24}{\lambda^{5/2}}\left(\frac{\pi ^2 \zeta_3}{2}-6 \zeta_5\right) + O(\lambda^{-3})\right] \, , \\[0.5em]
\mathcal{Y}_{1,2} \, &\underset{\lambda \rightarrow \infty}{\sim}\, N^2\left[1-\frac{8\pi^2}{3\lambda} +\frac{8}{\lambda^{3/2}}\left(\frac{2\pi^2}{3}+2\zeta_3\right)-\frac{8}{\lambda^{5/2}}\left(\frac{\pi^2\zeta_3}{2}+6\zeta_5\right) + O(\lambda^{-3})\right]\, \ .
\end{align}
\end{subequations}

\subsection{The next-to-planar term}
\label{subsec:JJJJNLO}

It is straightforward to check that also the next-to-planar term of the integrated correlators $\mathcal{Y}_f$ and $\mathcal{Y}_{1,2}$ deviate from the $\mathcal{N}=4$ SYM results of \cite{Chester:2020dja,Alday:2023pet} even at the perturbative level. Therefore, from now on, we exclusively focus on their expansions at large-$\lambda$. Given the complexity of this computation, we choose to compute only the leading term. Furthermore, it is advantageous to analyse each of the three contributions from \eqref{MConnectedFull}, as well as the contribution stemming from \eqref{HvevFull}, separately.

\subsubsection{Strong-coupling evaluation of \texorpdfstring{$\langle \mathcal{M}^{(1)\,2}_{\mathbb{Z}_2} \rangle_{\text{con}}$}{}}
As a first step, we use the identity \eqref{idM11} to show that the next-to-planar term of  $\langle \mathcal{M}^{(1)\,2}_{\mathbb{Z}_2} \rangle_{\text{con}}$ can be rewritten as 
\begin{align}
& 2\sum_{k=1}^{\infty}\sum_{\ell=1}^{\infty}\textsf{M}_{0,2k}\textsf{M}_{0,2\ell}\sqrt{k\,\ell}\left[\frac{(k^2+\ell^2-1)(k^2+\ell^2-14)}{12} -(k^2+\ell^2-1)\lambda\partial_{\lambda}\mathcal{F} - (\lambda\partial_\lambda)^2\mathcal{F}\right] \nonumber \\[0.5em]
& + \frac{\lambda}{8\pi^2}\sum_{k=1}^{\infty}\textsf{M}_{0,2k}\textsf{Q}_{0,2k} - \frac{\sqrt{2}\,\lambda}{96\pi^2}\sum_{k=1}^{\infty}\textsf{M}_{0,2k}\textsf{Q}_{2,2k} \,  .
\label{M1conNextToLeading}
\end{align}
The leading term of the strong-coupling expansions of the two contributions present in the second line of \eqref{M1conNextToLeading} can be evaluated by exploiting the expressions of the matrices $\textsf{M}_{0,2k}$, $\textsf{Q}_{0,2k}$, and $\textsf{Q}_{2,2k}$ as given in \eqref{Mmatrix}-\eqref{Qmatrix} and then employing the technique outlined in Appendix A of \cite{Alday:2023pet}. In this way we obtain
\begin{subequations}
\begin{align}
\frac{\lambda}{8\pi^2}\sum_{k=1}^{\infty}\textsf{M}_{0,2k}\textsf{Q}_{0,2k} \, & \underset{\lambda \rightarrow \infty}{\sim}\, \frac{1}{4} + O(\lambda^{-1/2})\, \\[0.5em] - \frac{\sqrt{2}\,\lambda}{96\pi^2}\sum_{k=1}^{\infty}\textsf{M}_{0,2k}\textsf{Q}_{2,2k} \, &\underset{\lambda \rightarrow \infty}{\sim}\, \frac{\sqrt{\lambda}}{144} + O(\lambda^{0}) \,  .
\end{align}
\label{M1QQ}
\end{subequations}
On the other hand, the first line of \eqref{M1conNextToLeading} requires more attention. We find it useful to expand the terms inside the square brackets. In this manner, we obtain a fourth-degree polynomial in  $k$ and $\ell$, allowing us to separately consider each distinct monomial. Specifically, the sums over the monomials of degree zero can be performed simply by using the identity \eqref{idM11} and read
\begin{align}
2\sum_{k=1}^{\infty}\sum_{\ell=1}^{\infty}\textsf{M}_{0,2k}\textsf{M}_{0,2\ell}\sqrt{k\ell}\left(\frac{7}{6}+\lambda\partial_{\lambda}\mathcal{F}-(\lambda\partial_{\lambda})^2\mathcal{F}\right) = \textsf{M}_{1,1}^2\left(\frac{7}{6}+\lambda\partial_{\lambda}\mathcal{F}-(\lambda\partial_{\lambda})^2\mathcal{F}\right)\,  .
\label{monomials0}
\end{align}
The summations over the monomials of degree two, instead, can be computed using the identity \eqref{idM11SECOND}, resulting in
\begin{align}
-2\sum_{k=1}^{\infty}\sum_{\ell=1}^{\infty}\textsf{M}_{0,2k}\textsf{M}_{0,2\ell}\sqrt{k\,\ell}\,(k^2+\ell^2)\left(\frac{5}{4}+\lambda\partial_{\lambda}\mathcal{F}\right) = -\left(\frac{5}{4}+\lambda\partial_{\lambda}\mathcal{F}\right)\left(1+\frac{1}{2}\lambda\partial_{\lambda}\right)\textsf{M}_{1,1}^2 \, .
\label{monomials2}
\end{align}
Finally, the summations over the monomials of degree four can be achieved by using the identities \eqref{idM11FIRST} and \eqref{idM11THIRD}. They read
\begin{align}
&\frac{1}{6}\sum_{k,\ell=1}^{\infty}\sqrt{k\,\ell}\,\textsf{M}_{0,2k}\,\textsf{M}_{0,2\ell}\,(k^4+\ell^4+2k^2\ell^2)  = \frac{1}{24}\left[\textsf{M}_{1,1}^2+2\textsf{M}_{1,1}\lambda\partial_{\lambda}\textsf{M}_{1,1}+ (\lambda\partial_{\lambda}\textsf{M}_{1,1})^2\right]
\nonumber \\
& -\frac{1}{6} \textsf{M}_{1,1}\frac{\sqrt{\lambda}}{8\pi^2} \int_0^{\infty}dt\, \frac{(t/2)^2}{\sinh(t/2)^2}J_1\left(\frac{t\sqrt{\lambda}}{2\pi}\right)\left[2\pi J_0\left(\frac{t\sqrt{\lambda}}{2\pi}\right)-t\sqrt{\lambda}J_1\left(\frac{t\sqrt{\lambda}}{2\pi}\right)\right] \,  .
\end{align}
Its strong-coupling expansion can be computed using the Mellin-Barnes method outlined in Section \ref{sec:2massder} and recalling \eqref{Mstrongcoupling}. This yields
\begin{align}
\frac{1}{6}\sum_{k,\ell=1}^{\infty}\sqrt{k\,\ell}\,\textsf{M}_{0,2k}\,\textsf{M}_{0,2\ell}\,(k^4+\ell^4+2k^2\ell^2) \, \underset{\lambda \rightarrow \infty}{\sim} \, -\frac{\sqrt{\lambda}}{144} + O(\lambda^{0})\, 
.
\label{monomials4}
\end{align}

We finally recollect that the strong-coupling expansion of the free energy of the $\mathbb{Z}_2$ quiver gauge theory has been computed in \cite{Beccaria:2022ypy}, where it was found that
\begin{align}
&\mathcal{F}(\lambda) \, \underset{\lambda \rightarrow \infty}{\sim} \, \frac{\sqrt{\lambda}}{4}+ O(\lambda^{0})\, . \label{DF} 
\end{align}
This way we can obtain the strong-coupling limits of the expressions \eqref{monomials0} and \eqref{monomials2}, namely
\begin{align}
& \textsf{M}_{1,1}^2\left(\frac{7}{6}+\lambda\partial_{\lambda}\mathcal{F}-(\lambda\partial_{\lambda})^2\mathcal{F}\right)\underset{\lambda \rightarrow \infty}{\sim} \frac{\sqrt{\lambda}}{64} + O\left( \lambda^0\right)\,, \\ 
& -\left(\frac{5}{4}+\lambda\partial_{\lambda}\mathcal{F}\right)\left(1+\frac{1}{2}\lambda\partial_{\lambda}\right)\textsf{M}_{1,1}^2  \underset{\lambda \rightarrow \infty}{\sim} -\frac{\sqrt{\lambda}}{32} + O\left( \lambda^0\right)\,.
\end{align}
Then, by including also the contributions from \eqref{M1QQ} and \eqref{monomials4}, we  determine the next-to-planar strong-coupling behaviour of $\langle \mathcal{M}^{(1)\,2}_{\mathbb{Z}_2}\rangle_{\text{con}}$, which reads 
\begin{align}
\langle \mathcal{M}^{(1)\,2}_{\mathbb{Z}_2} \rangle_{\text{con}} \, \underset{\lambda \rightarrow \infty}{\sim} \,   -\frac{\sqrt{\lambda}}{64} + O(\lambda^0) \,  .
\label{M1conStrong}
\end{align}

\subsubsection{Strong-coupling evaluation of \texorpdfstring{$\langle \mathcal{M}^{(2)\,2}_{\mathbb{Z}_2}\rangle_{\text{con}}$}{}}
\label{sec:strongM22con}
Let us begin by examining the first contribution present in the r.h.s of \eqref{M2connected}. The computation of the leading term of its strong-coupling expansion has been carried out in Appendix \ref{app:TrM2strong}. Here, we simply present the result, which is as follows
\begin{align}
\frac{1}{2}\,\left(\text{Tr}[(\textsf{M}^{\text{even}})^2]+\text{Tr}[(\textsf{M}^{\text{odd}})^2]\right) \, \underset{\lambda \rightarrow \infty}{\sim} \, \frac{\sqrt{\lambda}}{4}\left(\frac{1}{3}-\frac{\pi ^2}{45}\right) + O(\lambda^{0})\, \ .
\label{M2first}
\end{align}
On the other hand, the strong-coupling evaluation of the other two terms present in \eqref{M2connected} requires more attention. We find it useful to rewrite this contribution as
\begin{align}
\frac{1}{2}\left(\text{Tr}[\textsf{M}^{\text{even}}\,\textsf{D}^{\text{even}}\,\textsf{M}^{\text{even}}\,\textsf{D}^{\text{even}}]+\text{Tr}[\textsf{M}^{\text{odd}}\,\textsf{D}^{\text{odd}}\,\textsf{M}^{\text{odd}}\,\textsf{D}^{\text{odd}}]\right) = \frac{1}{2}\left(\mathcal{I}^{(1)}+2\,\mathcal{I}^{(2)}+ \mathcal{I}^{(3)}\right)\, , 
\label{M2second}
\end{align}
with 
\begin{align}
\mathcal{I}^{(2)} = \sum_{k_1=1}^{\infty}\mathcal{I}_{k_1}^{(2)}\, , \qquad \mathcal{I}^{(3)} = \sum_{k_1,k_2=1}^{\infty}\mathcal{I}_{k_1,k_2}^{(3)}
\label{I2andI3}
\end{align}
and where
\begin{subequations}
\begin{align}
&\mathcal{I}^{(1)} =  \sum_{q_1,p_1=1}^{\infty}\,\left[\textsf{M}_{p_1,q_1}^{\text{even}}\textsf{M}_{q_1,p_1}^{\text{even}} + \textsf{M}_{p_1,q_1}^{\text{odd}}\textsf{M}_{q_1,p_1}^{\text{odd}}\right] = \text{Tr}[(\textsf{M}^{\text{even}})^2]+\text{Tr}[(\textsf{M}^{\text{odd}})^2] \,  , \label{con1}\\[0.5em]
&\mathcal{I}^{(2)}_{k_1} = \sum_{q_1,p_1=1}^{\infty}\sum_{q_2,p_2=1}^{\infty}\,\left[\textsf{M}^{\text{even}}_{p_1,q_1}\delta_{q_1,p_2}\textsf{M}^{\text{even}}_{p_2,q_2}(\textsf{X}^{\text{even}})_{q_2,p_1}^{k_1} + \textsf{M}^{\text{odd}}_{p_1,q_1}\delta_{q_1,p_2}\textsf{M}^{\text{odd}}_{p_2,q_2}(\textsf{X}^{\text{odd}})_{q_2,p_1}^{k_1}\right]\, , \label{con2}\\[0.5em] &\mathcal{I}^{(3)}_{k_1,k_2} = \sum_{q_1,p_1=1}^{\infty}\sum_{q_2,p_2=1}^{\infty}\left[\,\textsf{M}^{\text{even}}_{p_1,q_1}(\textsf{X}^{\text{even}})_{q_1,p_2}^{k_1}\textsf{M}^{\text{even}}_{p_2,q_2}(\textsf{X}^{\text{even}})_{q_2,p_1}^{k_2} + \textsf{M}^{\text{odd}}_{p_1,q_1}(\textsf{X}^{\text{odd}})_{q_1,p_2}^{k_1}\textsf{M}^{\text{odd}}_{p_2,q_2}(\textsf{X}^{\text{odd}})_{q_2,p_1}^{k_2}\right] \, , \label{con3}
\end{align}
\end{subequations}
with $k_1,k_2 \geq 1$. We observe that the leading term of the strong-coupling expansions of $\mathcal{I}^{(1)}$ is simply given by \eqref{resultI1}. Instead, the corresponding computation for  $\mathcal{I}^{(2)}$ and $\mathcal{I}^{(3)}$ is quite long and has been detailed in Appendix \ref{app:StrongCoupling}. Here, we simply report the results, which are 
\begin{subequations}
\begin{align}
& \mathcal{I}^{(2)} \underset{\lambda \rightarrow \infty}{\sim} \,\frac{\sqrt{\lambda}}{2}\left(\frac{19 \pi ^2}{720}-\frac{1}{3}\right)  +O(\lambda^{0})\, ,  \\[0.5em]
& \mathcal{I}^{(3)}  \, \underset{\lambda \rightarrow \infty}{\sim} \, \frac{\sqrt{\lambda}}{2}\left(\frac{1}{6}-\frac{\pi^2}{90}\right) + O(\lambda^0)\,  .   
\end{align}
\label{IIstrong}
\end{subequations}
Thus, the leading term of the strong-coupling expansion of $\langle \mathcal{M}^{(2)\,2}_{\mathbb{Z}_2}\rangle_{\text{con}}$ can be easily obtained by summing up \eqref{M2first} and \eqref{M2second} using the expressions \eqref{IIstrong}. This yields
\begin{align}
\langle \mathcal{M}^{(2)\,2}_{\mathbb{Z}_2}\rangle_{\text{con}} \, \underset{\lambda \rightarrow \infty}{\sim} \, \sqrt{\lambda}\left(\frac{1}{24}-\frac{\pi ^2}{1440}\right)+ O(\lambda^0)\, .  
\label{M2conSTRONG}
\end{align}

\subsubsection{Strong-coupling evaluation of \texorpdfstring{$2\langle \mathcal{M}^{(1)}_{\mathbb{Z}_2}\mathcal{M}^{(2)}_{\mathbb{Z}_2}\, \  \rangle_{\text{con}}$}{}}
We begin by examining the contribution in the second line of \eqref{M1M2con}. Its large-$\lambda$ expansion can be computed using the expression of the $\textsf{M}$ matrix given in \eqref{Mmatrix},  the free energy expansion \eqref{DF}, and employing the techniques outlined in Appendix A of \cite{Alday:2023pet}. This way, we analytically obtain
\begin{align}
-2\,(\lambda\partial_{\lambda}\mathcal{F})\,\sum_{k=1}^{\infty}\sum_{p=1}^{\infty}\textsf{M}_{0,2k}\,\textsf{M}_{2k,2p}\,\sqrt{2p} \, \underset{\lambda \rightarrow \infty}{\sim} \, \frac{\sqrt{\lambda}}{32} + O(\lambda^0)\,  .
\label{M2M1c1}
\end{align}
On the other hand, evaluating the contribution in the first line of \eqref{M1M2con} for large $\lambda$ is more intricate due to the series over $\textsf{d}_{p}\,\textsf{d}_q$ coefficients and we are not aware of a procedure that would allow us to perform such computation analytically. Therefore, we choose to employ numerical techniques. A comprehensive discussion of this analysis is provided in Appendix \ref{app:numerics}; here, we simply report the corresponding result, which reads\,\footnote{A more accurate numerical estimate of this quantity is provided in \cite{DeSmet:2025mbc}, yielding a value of $-\frac{\sqrt{\lambda}}{48}$ with a precision of up to $12$ significant digits.} 
\begin{align}
-\textsf{M}_{1,1}\sum_{p,q=2}^{\infty}(-1)^{p-pq}\,\textsf{M}_{p,q}\,(\sqrt{p\,q}-\textsf{d}_p\textsf{d}_q) \, \underset{\lambda \rightarrow \infty}{\sim} \, -\frac{\pi^2\sqrt{\lambda}}{480} + O(\lambda^0)\, .
\label{M2M1c2}
\end{align}
Therefore, by combining the expressions \eqref{M2M1c1} and \eqref{M2M1c2}, we obtain
\begin{align}
2\langle \mathcal{M}^{(1)}_{\mathbb{Z}_2}\mathcal{M}^{(2)}_{\mathbb{Z}_2}\, \  \rangle_{\text{con}}  \, \underset{\lambda \rightarrow \infty}{\sim} \, \sqrt{\lambda}\left(\frac{1}{32} - \frac{\pi^2}{480}\right) + O(\lambda^0) \,  .
\label{M1M2conSTRONG}
\end{align}

\subsubsection{Strong-coupling evaluation of \texorpdfstring{$\langle \mathcal{H}_{\mathbb{Z}_2} \rangle$}{}}
Let us start by examining the second and third terms in the first line of \eqref{HvevFull}. By using the identities \eqref{id:Q12} and \eqref{id:Q11}, together with the strong-coupling expansion of the free energy \eqref{DF} and employing the Mellin-Barnes method, one can demonstrate that
\begin{subequations}
\begin{align}
&- \frac{1}{12}\left(\textsf{Q}_{1,1}-\frac{1}{6}\sum_{k=1}^{\infty}\sqrt{2k+1}\textsf{Q}_{1,2k+1}\right) \, \underset{\lambda \rightarrow \infty}{\sim} \,  O(\lambda^{-1/2})\, \ , \\[0.5em]
& -\frac{N}{6\sqrt{2}}\sum_{k=1}^{\infty}\textsf{Q}_{0,2k}\,\langle \mathcal{P}_{2k}^{+} \rangle \simeq -(\lambda\partial_{\lambda}\mathcal{F})\textsf{Q}_{1,1} \, \underset{\lambda \rightarrow \infty}{\sim} \, \frac{\pi^2}{144} + O(\lambda^{-1/2})\, \ .
\end{align}
\label{H2first}
\end{subequations}
On the other hand the strong-coupling evaluation of the term in the second line of \eqref{HvevFull} is more subtle and its computation has been detailed in Appendix \ref{app:StrongCouplingQD}. Here, we simply report the result, namely
\begin{align}
-\frac{1}{24}\sum_{p,q=2}^{\infty}(-1)^{p-pq}\,\textsf{Q}_{p,q}\,(\delta_{p,q}-\textsf{D}_{p,q}) = -\frac{1}{24}\text{Tr}[\textsf{Q}(\mathbb{1}-\textsf{D})] \,  \underset{\lambda \rightarrow \infty}{\sim} \, \frac{\pi^2\sqrt{\lambda}}{1440} + O(\lambda^0)\, .
\label{H2second}
\end{align}
Therefore, combining \eqref{H2first} and \eqref{H2second}, we conclude that at strong-coupling the next-to-planar contribution arising from $\langle \mathcal{H}_{\mathbb{Z}_2} \rangle$ is given by
\begin{align}
\langle \mathcal{H}_{\mathbb{Z}_2} \rangle     \,  \underset{\lambda \rightarrow \infty}{\sim} \, \frac{\pi^2\sqrt{\lambda}}{1440} + O(\lambda^0)\,  . 
\label{H2STRONG}
\end{align}

\subsubsection{Next-to-planar term final result}

We insert the expressions \eqref{M1conStrong}, \eqref{M2conSTRONG}, \eqref{M1M2conSTRONG} and \eqref{H2STRONG} in \eqref{JJJJ} and this yields the next-to-planar contribution of the integrated correlators \eqref{JJJJ} at strong-coupling
\begin{subequations}
\begin{align}
\mathcal{Y}_f \, & \underset{\lambda \rightarrow \infty}{\sim} \,   \sqrt{\lambda}\left(\frac{11}{16}-\frac{\pi ^2}{60}\right) + O(\lambda^0)\,  , \\
\mathcal{Y}_{1,2} \, & \underset{\lambda \rightarrow \infty}{\sim}\, \sqrt{\lambda}\left(\frac{11}{48}-\frac{\pi^2}{90}\right) + O(\lambda^0)\, \ .
\end{align}
\label{JJJJNLO}
\end{subequations}

This is our final result for these observables. An examination of expression (2.23) in \cite{Alday:2023pet} reveals that the corresponding $\mathcal{N}=4$ integrated correlator scales with the same power of the 't Hooft coupling, namely $\sqrt{\lambda}$.

\section{Conclusions}
\label{conclusion}

By employing matrix model techniques, we investigated two distinct types of integrated correlators in the large-$N$ limit of the $\mathcal{N}=2$ quiver gauge theory arising as a $\mathbb{Z}_2$ orbifold of $\mathcal{N}=4$ SYM. The first kind  of integrated correlator \eqref{Z2intcorr} involves the insertion of two moment map operators and two untwisted or twisted chiral operators with conformal dimension $p$, respectively $\mathcal{C}^{+}_{f,p}$ and $\mathcal{C}^{-}_{f,p}$. Specifically, we analytically derived the leading term of the large-$N$ expansion of this correlator, followed by an examination of its behaviour in the strong-coupling regime, which led to the following result
\begin{align}
\lim_{\lambda \rightarrow \infty }\frac{\mathcal{C}_{f,p}^+}{\mathcal{G}_p^{(0)}} \,=\, 
\lim_{\lambda \rightarrow \infty }\frac{\mathcal{C}^-_{f,p}}{\mathcal{G}_p} \, =\, \frac{p-1}{4} \, ,
\end{align}
where $\mathcal{G}_p^{(0)}$ and $\mathcal{G}_p$ denote the 2-point function between the untwisted and twisted operators, respectively. Remarkably, the same behaviour at strong coupling, up to a numerical factor, has been previously observed for both $\mathcal{N}=4$ SYM \cite{Binder:2019jwn} and the $\mathcal{N}=2$ $\textbf{E}$-theory analysed in \cite{Billo:2023kak}. Consequently, it would be interesting to explore in future studies the extent to which this behaviour may be universal by evaluating the integrated correlators \eqref{Z2intcorr} in other $\mathcal{N}=2$ SCFTs. 
Furthermore, it would also be intriguing to go beyond the planar term of the large-$N$ expansion of these correlators. This, in turn, would necessitate deriving the expressions for 3- and 4-point functions involving $\mathcal{P}_{k}^{\pm}$ operators analogous to \eqref{PPplus}. We choose to leave it for future investigation.

The second type of integrated correlators \eqref{4mY} involves the insertion of 4 moment map operators and it has been analysed in Section \ref{sec:JJJJ}. For these correlators we derived exact expressions that hold for any value of the 't Hooft coupling at both the planar and next-to-planar orders of their large-$N$ expansion. Finally we considered the strong coupling limit of these correlators and we found
\begin{subequations}
\begin{align}
& \mathcal{Y}_f  \, \underset{\lambda \rightarrow \infty}{\sim}\, N^2\left[3+\frac{8\pi^2}{\lambda}+\frac{24}{\lambda^{3/2}}\left(2\zeta_3-\frac{2 \pi ^2}{3}\right)+O(\lambda^{-5/2})\right]   +\left[\sqrt{\lambda}\left(\frac{11}{16}-\frac{\pi ^2}{60}\right) + O(\lambda^0)\right] + O(N^{-2})\, , \\[0.5em]
& \mathcal{Y}_{1,2} \, \underset{\lambda \rightarrow \infty}{\sim}\, N^2\left[1-\frac{8\pi^2}{3\lambda} +\frac{8}{\lambda^{3/2}}\left(\frac{2\pi^2}{3}+2\zeta_3\right)+ O(\lambda^{-5/2})\right]+ \left[\sqrt{\lambda}\left(\frac{11}{48}-\frac{\pi^2}{90}\right) + O(\lambda^0)\right] + O(N^{-2})\,  .  
\end{align}
\label{FinalResultJJJJ}  
\end{subequations}
In the future, we would like to investigate whether the subleading terms at the next-to-planar order of the expansions \eqref{FinalResultJJJJ} can  be computed by generalising the method developed in \cite{Belitsky:2020qir,Belitsky:2020qrm}.
Furthermore, the $\mathbb{Z}_2$ quiver gauge theory has a known gravity dual geometry, %(see for instance \cite{Kachru:1998ys,Gukov:1998kk}). 
consequently, the quantum field theory  predictions presented in this article can be confirmed through the AdS/CFT correspondence, generalizing the methodology outlined in \cite{Binder:2019jwn,Chester:2020dja} in the case of $\mathcal{N}=4$ SYM.

As a final remark, it is worth noting that the methods developed in this work can be extended to the case of generic quiver gauge theories with $M$ nodes. It would be interesting to analyse the integrated correlators in this broader context and explore how the results presented in this paper might be influenced by the different orbifold projection.

\vskip 1cm
\noindent {\large {\bf Acknowledgments}}
\vskip 0.2cm
We are very grateful to M. Billò, M. Frau and A. Lerda for many important discussions and for reading and commenting on the draft of our article. Furthermore we would like to thank P.-J. De Smet, F. Galvagno, H. Paul, D. Rodr\'iguez-G\'omez, K. Zarembo and X. Zhang for useful discussions. In particular we are grateful to P.-J. De Smet for pointing out an alternative numerical method different from the one illustrated in Appendix \ref{app:numerics}. Moreover, PV would like to thank the Quantum Field and String Theory Group of Humboldt University for the hospitality during the first stages of this project.
The work of PV is partially supported by the MUR PRIN contract 2020KR4KN2 ``String Theory as a bridge between Gauge Theories and Quantum Gravity'' and by the INFN project ST\&FI ``String Theory \& Fundamental Interactions''. The work of AP is supported  by the Deutsche Forschungsgemeinschaft (DFG, German Research Foundation) via the Emmy Noether program ``Exploring the landscape of string theory flux vacua using exceptional field theory” (project number 426510644). 

\vskip 1cm

\appendix
\section{Correlation functions among \texorpdfstring{$\mathcal{P}_{n}^{\pm}$}{} operators}
\label{app:Pcorrelators}
In this appendix, since they are extensively used for computing the integrated correlators, we find it useful to gather the expressions for the correlation functions (up to 4-point) among the $\mathcal{P}_{n}^{\pm}$ operators \eqref{PP} in the large-$N$ limit of both the free Gaussian matrix model and the interacting $\mathbb{Z}_2$ quiver gauge theory. These properties have been mainly studied in \cite{Billo:2021rdb,Billo:2022fnb}, here we just recollect the results. Let us start by recalling that the  relation \eqref{PP} can be inverted, allowing us to express the $\mathcal{P}_{n}^{\pm}$ operators as functions of the $A_{n}^{\pm}$ operators \eqref{AA}
\begin{align}
\label{Pn}
\mathcal{P}^{\pm}_n = \sqrt{n}\left(\frac{2}{N}\right)^{\frac{n}{2}}\left[\sum_{k=0}^{\lfloor \frac{n-1}{2} \rfloor}\left(-\frac{N}{2}\right)^k \frac{(n-k-1)!}{(n-2k)!\,k!}\left(A^{\pm}_{n-2k} - \langle A^{\pm}_{n-2k}\rangle_0\right) \right]  \, .
\end{align}

\subsection{Gaussian matrix model}
The non-trivial (up to 3-point) correlation functions in the free theory are as follows

\begin{equation}
\langle \mathcal{P}_{n}^{\pm} \rangle_0 = 0\, , \ \  \  \ \ \langle \mathcal{P}^{\pm}_{n_1}\,\mathcal{P}^{\pm}_{n_2} \rangle_0 = \delta_{n_1,n_2}\,, \, \ \ \  \langle \mathcal{P}^{\pm}_{n_1}\,\mathcal{P}^{\pm}_{n_2}\,\mathcal{P}^{\pm}_{n_3} \rangle_0 \simeq \frac{\sqrt{n_1\,n_2\,n_3} }{\sqrt{2}N}\,  . 
\end{equation} 
As discussed in \cite{Billo:2021rdb} the 4-point functions can be computed using Wick's theorem, namely
\begin{align}
& \langle \mathcal{P}_{n_1}^{\pm}\,\mathcal{P}_{n_2}^{\pm}\,\mathcal{P}_{n_3}^{\pm}\,\mathcal{P}_{n_4}^{\pm} \rangle_0 \simeq \langle \mathcal{P}_{n_1}^{\pm}\,\mathcal{P}_{n_2}^{\pm} \rangle_0 \,\langle \mathcal{P}_{n_3}^{\pm}\,\mathcal{P}_{n_4}^{\pm} \rangle_0 + \langle \mathcal{P}_{n_1}^{\pm}\,\mathcal{P}_{n_3}^{\pm} \rangle_0 \, \langle \mathcal{P}_{n_2}^{\pm}\,\mathcal{P}_{n_4}^{\pm} \rangle_0 
 +\langle \mathcal{P}_{n_1}^{\pm}\,\mathcal{P}_{n_4}^{\pm} \rangle_0 \,\langle \mathcal{P}_{n_2}^{\pm}\mathcal{P}_{n_3}^{\pm} \rangle_0 \, , \\
& \langle \mathcal{P}^{+}_{n_1}\,\mathcal{P}^{+}_{n_2}\,\mathcal{P}^{-}_{n_3}\,\mathcal{P}^{-}_{n_4} \rangle_0 \simeq \langle \mathcal{P}_{n_1}^{+}\mathcal{P}_{n_2}^{+} \rangle_0\, \langle \mathcal{P}_{n_3}^{-}\mathcal{P}_{n_4}^{-} \rangle_0 \, \ .  
\end{align}
\subsection{Interacting \texorpdfstring{$\mathbb{Z}_2$}{} quiver gauge theory}
The non-trivial (up to 4-point) correlation functions in the interacting theory are as follows.
\paragraph{1-point function}
The 1-point function of an even untwisted operator exhibits non-trivial dependence on the 't Hooft coupling and can be expressed as a function of the free energy in the following manner
\begin{align}
\label{1pointPLUS}
\langle \mathcal{P}_{2n}^{+} \rangle \, \simeq \, - \frac{\sqrt{n}}{N}\,\lambda \partial_{\lambda}\mathcal{F}\, \ .   
\end{align}
To prove this last relation, we use the following large-$N$ expansion
\begin{align}
\label{vevAeven}
\langle A_{2n}^{+} \rangle \, \simeq \, \langle A_{2n}^{+} \rangle_0\left[1-\frac{n(n+1)}{2N^2}\lambda\partial_{\lambda}\mathcal{F}+ O\left(\frac{1}{N^4}\right)\right]\, ,
\end{align}
together with
\begin{align}
& \langle A_{2n}^{+} \rangle_0 \, \simeq \, \sqrt{2}\,\frac{N^{n+1}}{2^{n}}\frac{(2n)!}{(n+1)!\,n!} + O(N^{n-1})\, .
\end{align}
We then substitute expression \eqref{vevAeven} into \eqref{Pn} and after some algebraic manipulations we recover the expression \eqref{1pointPLUS}, namely
\begin{align}
& \langle \mathcal{P}_{2n}^{+} \rangle \, \simeq \, -\frac{\sqrt{n}}{N}\left(\lambda\partial_{\lambda}\mathcal{F}\right)\sum_{i=0}^{n}\frac{(-1)^{i}(2n-i-1)!}{i!\,(n-i)!\,(n-i-1)!} \, = \, - \frac{\sqrt{n}}{N}\,\lambda\partial_{\lambda}\mathcal{F} \, . \label{1pPeven}
\end{align}
\paragraph{2-point functions}
\begin{align}
\label{2point2}
\langle \mathcal{P}^{+}_{n_1}\,\mathcal{P}^{+}_{n_2} \rangle \simeq \delta_{n_1,n_2}\, , \ \ \ \ \langle \mathcal{P}^{-}_{n_1}\,\mathcal{P}^{-}_{n_2} \rangle \simeq \textsf{D}_{n_1,n_2}\, \ .
\end{align}
\paragraph{3-point functions} 
\begin{subequations}
\begin{align}
& \langle \mathcal{P}^{+}_{n_1}\,\mathcal{P}^{+}_{n_2}\,\mathcal{P}^{+}_{n_3} \rangle \, \simeq \,  \frac{\sqrt{n_1\,n_2\,n_3}}{\sqrt{2}N} + \langle \mathcal{P}^{+}_{n_1} \rangle \,\delta_{n_2,n_3} + \langle \mathcal{P}^{+}_{n_2} \rangle \, \delta_{n_1,n_3} + \langle \mathcal{P}^{+}_{n_3} \rangle \, \delta_{n_1,n_2} \, \ ,\\[0.5em]
& \langle \mathcal{P}_{n_1}^{+}\,\mathcal{P}_{n_2}^{-}\,\mathcal{P}_{n_3}^{-} \rangle \, \simeq \, \frac{\sqrt{n_1}\,\textsf{d}_{n_2}\,\textsf{d}_{n_3}}{\sqrt{2}N} +\langle \mathcal{P}_{n_2}^{-}\,\mathcal{P}_{n_3}^{-} \rangle \, \langle \mathcal{P}_{n_1}^{+} \rangle \, \ ,
\end{align}
\end{subequations}
where $\textsf{D}_{n,m}$ and $\textsf{d}_n$ are defined in \eqref{dandD}.
\paragraph{4-point functions} As demonstrated in \cite{Billo:2021rdb}, the 4-point functions can be factorised using Wick's theorem as follows
\begin{subequations}
\label{4point2}
\begin{align}
& \langle \mathcal{P}^{+}_{n_1}\,\mathcal{P}^{+}_{n_2}\,\mathcal{P}^{+}_{n_3}\,\mathcal{P}^{+}_{n_4} \rangle \, \simeq \, \delta_{n_1,n_2}\,\delta_{n_3,n_4}+ \delta_{n_1,n_3}\,\delta_{n_2,n_4}+\delta_{n_1,n_4}\,\delta_{n_2,n_3} \,  , \\[0.5em]
& \langle \mathcal{P}^{-}_{n_1}\,\mathcal{P}^{-}_{n_2}\,\mathcal{P}^{-}_{n_3}\,\mathcal{P}^{-}_{n_4} \rangle \, \simeq \, \textsf{D}_{n_1,n_2}\,\textsf{D}_{n_3,n_4}+ \textsf{D}_{n_1,n_3}\,\textsf{D}_{n_2,n_4}+\textsf{D}_{n_1,n_4}\,\textsf{D}_{n_2,n_3} \,  , \\[0.5em]
& \langle \mathcal{P}^{+}_{n_1}\,\mathcal{P}^{+}_{n_2}\,\mathcal{P}^{-}_{n_3}\,\mathcal{P}^{-}_{n_4} \rangle \, \simeq \,  \langle \mathcal{P}^{+}_{n_1}\,\mathcal{P}^{+}_{n_2}  \rangle \, \langle \mathcal{P}^{-}_{n_3}\,\mathcal{P}^{-}_{n_4}\rangle\, \ = \delta_{n_1,n_2}\,\textsf{D}_{n_3,n_4}  .   
\end{align}
\end{subequations}

\section{Details on the large-\texorpdfstring{$N$}{} expressions of \texorpdfstring{$\mathcal{M}_{\mathbb{Z}_2}$}{} and \texorpdfstring{$\mathcal{H}_{\mathbb{Z}_2}$}{}}
\label{app:MandH}

In this appendix we give details about the large-$N$ expansions \eqref{M2full}-\eqref{H2full} of the operators $\mathcal{M}_{\mathbb{Z}_2}$ and $\mathcal{H}_{\mathbb{Z}_2}$, which are exact in the 't Hooft coupling $\lambda$. Firstly, using the untwisted/twisted operator basis defined in \eqref{AA}, $\mathcal{M}_{\mathbb{Z}_2}$ and $\mathcal{H}_{\mathbb{Z}_2}$ can be rewritten in the following form
\begin{align}
&  \mathcal{M}_{\mathbb{Z}_2} =-\frac{1}{2} \sum_{n=1}^{\infty}\sum_{\ell=0}^{2n}(-1)^{n+\ell}\,\frac{(2n+1)!\,\zeta_{2n+1}}{(2n-\ell)!\,\ell!}\left(\frac{\lambda}{8\pi^2N}\right)^n \left(A^+_\ell A^+_{2n-\ell}-A^+_\ell A^-_{2n-\ell}+A^-_\ell A^+_{2n-\ell}-A^-_\ell A^-_{2n-\ell} \right) , \label{Mstep2}
\\[0.5em]
& \mathcal{H}_{\mathbb{Z}_2} =  - \frac{1}{24}\sum_{n=1}^{\infty}\sum_{\ell=0}^{2n-2}(-1)^{n+\ell}\,\frac{(2n+1)!\,\zeta_{2n+1}}{(2n-2-\ell)!\,\ell!}\left(\frac{\lambda}{8\pi^2N}\right)^{n-1}\left(A^+_\ell A^+_{2n-\ell}-A^+_\ell A^-_{2n-\ell}+A^-_\ell A^+_{2n-\ell}-A^-_\ell A^-_{2n-\ell} \right)  .
\label{Hstep2}
\end{align}
Then we perform the change of basis $A^\pm_k \mapsto \mathcal{P}_k^\pm$ introduced in \eqref{PP}. At this stage it is convenient to recollect that the one-point functions of the $A_k^{\pm}$ operators in the free matrix model, corresponding to the $\mathbb{Z}_2$ quiver, read \cite{Billo:2021rdb}
\begin{align}
\label{1ptA}
\langle A_k^+ \rangle_0 = \sqrt{2}\,t_k \ ,\ \ \ \ \ \ \ \ \langle A_k^- \rangle_0 = 0
\end{align}
where the $t_k$ have been computed by employing the expectation value in the Gaussian matrix model as defined in \eqref{tt}. In particular the VEV of even traces in the large-$N$ limit is given by
\begin{align}
\label{eventraces}
t_{2k} \, = \, \frac{N^{k+1}}{2^k} \frac{(2k)!}{k!(k+1)!} - \frac{N^{k-1}}{2^{k+1}} \frac{(2k)!}{k!(k-1)!} \left(1-\frac{k-1}{6} \right) + O(N^{k-3})\,.
\end{align}
Now we can proceed with the derivation of the expressions \eqref{M2full} and \eqref{H2full}.

\subsection{The operator \texorpdfstring{$\mathcal{M}_{\mathbb{Z}_2}$}{}}

Exploiting the expressions \eqref{1ptA} it is immediate to realize that the term with zero $\mathcal{P}^\pm_k$ operators, $\mathcal{M}_{\mathbb{Z}_2}^{(0)}$, is equal to the corresponding operator in the $\textbf{E}$-theory defined in \cite{Billo:2023kak}, i.e.
\begin{align}
& \mathcal{M}_{\mathbb{Z}_{2}}^{(0)} = -\sum_{n=1}^{\infty}\sum_{\ell=0}^{n}(-1)^{n}\frac{\zeta_{2n+1}(2n+1)!}{(2n-2\ell)!(2\ell)!}\left(\frac{\lambda}{8\pi^2N}\right)^n t_{2\ell}\,t_{2n-2\ell} =  \, \mathcal{M}_{\textbf{E}}^{(0)} \, .
\end{align}
Thus performing the same steps as in Appendix A of \cite{Billo:2023kak} one finds
\begin{align}
\mathcal{M}_{\mathbb{Z}_2}^{(0)} = \left[N^2\,\mathsf{M}_{0,0} + \mathsf{M}_{1,1} -\frac{1}{6}\sum_{k=1}^{\infty}\sqrt{2k+1}\mathsf{M}_{1,2k+1} + O\left(\frac{1}{N^2}\right)\right] \,.
\end{align}
Now we focus on the term with one $\mathcal{P}^\pm_k$ operator, $\mathcal{M}_{\mathbb{Z}_2}^{(1)}$. After some straightforward algebraic simplifications and using \eqref{1ptA}, we are left with
\begin{align}
\label{M1Z2I}
& \mathcal{M}_{\mathbb{Z}_{2}}^{(1)} = -\sqrt{2}\sum_{n=1}^{\infty}\sum_{\ell=0}^{n}\sum_{q=0}^{\ell}\sqrt{2\ell-2q}\,\frac{(-1)^{n}(2n+1)!\,\zeta_{2n+1}}{(2n-2\ell)!\,q!(2\ell-q)!}\left(\frac{\lambda}{8\pi^2N}\right)^n \left(\frac{N}{2}\right)^\ell t_{2n-2\ell}\,\mathcal{P}^+_{2\ell-2q}
\end{align}
which, recalling that the untwisted even operators of the 
$\mathbb{Z}_2$ quiver correspond to the even operators of the \textbf{E}-theory, i.e. $\mathcal{P}^{+,\,{\mathbb{Z}_2}}_{2k}\equiv \mathcal{P}^{\textbf{E}}_{2k}$, coincides with $\mathcal{M}_\textbf{E}^{(1)}$ up to the multiplicative constant $1/\sqrt{2}$. Therefore, for the leading term, we follow the analysis performed in Appendix A of \cite{Billo:2023kak} and we obtain the expression
\begin{align}
\mathcal{M}_{\mathbb{Z}_{2}}^{(1)} = \sqrt{2}N\sum_{k=1}^{\infty}\mathsf{M}_{0,2k}\mathcal{P}_{2k}^{+} + O(N^{-1})\,.
\end{align}
However here we want to find also the $O(N^{-1})$-term of the large-$N$ expansion of $\mathcal{M}_{\mathbb{Z}_{2}}^{(1)}$. In order to do that, using \eqref{eventraces} we plug in \eqref{M1Z2I} the subleading contribution of $t_{2n-2\ell}$, so that we end up with
\begin{align}
\frac{1}{\sqrt{2}\,N}\sum_{n=1}^{\infty}\sum_{\ell=0}^{n}\sum_{q=0}^{\ell}\sqrt{2\ell-2q}\,\frac{(-1)^{n}(2n+1)!\,\zeta_{2n+1}}{q!(2\ell-q)!(n-\ell)!(n-\ell-1)!}\left(\frac{\lambda}{16\pi^2}\right)^n \left(1-\frac{n-\ell-1}{6} \right)\mathcal{P}^+_{2\ell-2q} \,. 
\end{align}
Then we exploit the identity \eqref{integralidentity} and we rewrite the previous expression as
\begin{align}
& \frac{1}{\sqrt{2}\,N}\int_0^\infty \frac{dt}{t}\,\frac{(t/2)^2}{\sinh(t/2)^2}\,\sum_{n=1}^{\infty}\sum_{\ell=0}^{n}\sum_{q=0}^{\ell}\,\frac{(-1)^{n}\sqrt{2\ell-2q}\,\mathcal{P}^+_{2\ell-2q}}{q!\,(2\ell-q)!\,(n-\ell)!\,(n-\ell-1)!}\left(\frac{\sqrt{\lambda}t}{4\pi}\right)^{2n} \notag \\
& -\frac{1}{6\,\sqrt{2}\,N}\int_0^\infty \frac{dt}{t}\,\frac{(t/2)^2}{\sinh(t/2)^2}\,\sum_{n=1}^{\infty}\sum_{\ell=0}^{n}\sum_{q=0}^{\ell}\,\frac{(-1)^{n}\,\sqrt{2\ell-2q}\,\mathcal{P}^+_{2\ell-2q}}{q!\,(2\ell-q)!\,(n-\ell)!\,(n-\ell-2)!}\left(\frac{\sqrt{\lambda}t}{4\pi}\right)^{2n} \,.
\end{align}
Hence, relabelling the summation indices we can resum the series in $\lambda$ using the expansion of the Bessel functions of the first kind
\begin{align}
\label{besselJ}
J_n(x) = \sum_{k=0}^{\infty} \frac{(-1)^k}{k!(k+n)!}\left( \frac{x}{2} \right)^{2k+n} \,,
\end{align}
and we obtain
\begin{align}
& -\frac{\sqrt{2}\,\lambda}{32\pi^2 N}\int_0^{\infty}\,dt\, t\, \frac{(t/2)^2}{\sinh(t/2)^2}\sum_{k=1}^{\infty}(-1)^{k}\sqrt{2k}\left(\frac{4\pi}{t\sqrt{\lambda}}\right)J_1\left(\frac{t\sqrt{\lambda}}{2\pi}\right)J_{2k}\left(\frac{t\sqrt{\lambda}}{2\pi}\right)\mathcal{P}_{2k}^{+} \notag \\
& - \frac{\lambda}{192\pi^2 N}\int_0^{\infty}\,dt\, t\, \frac{(t/2)^2}{\sinh(t/2)^2}\sum_{k=1}^{\infty}(-1)^{k}2\sqrt{k}\,J_2\left(\frac{t\sqrt{\lambda}}{2\pi}\right)J_{2k}\left(\frac{t\sqrt{\lambda}}{2\pi}\right)\mathcal{P}_{2k}^{+} \,.
\end{align}
Exploiting the definition of the matrix elements in \eqref{Q0n}-\eqref{Qnm} we recover 
\begin{align}
\mathcal{M}_{\mathbb{Z}_2}^{(1)} = \sqrt{2}N\sum_{k=1}^{\infty}\mathsf{M}_{0,2k}\mathcal{P}_{2k}^{+} + \frac{\sqrt{2}\,\lambda}{32\pi^2 N}\sum_{k=1}^{\infty}\mathsf{Q}_{0,2k}\mathcal{P}_{2k}^{+} - \frac{\lambda}{192\pi^2 N}\sum_{k=1}^{\infty}\mathsf{Q}_{2,2k}\,\mathcal{P}_{2k}^{+} +O(N^{-3})\,.
\end{align}
Finally we consider the term $\mathcal{M}_{Z_2}^{(2)}$ proportional to two $\mathcal{P}^{\pm}_k$ operators that, after easy manipulations, is given by
\begin{align}
\mathcal{M}_{\mathbb{Z}_2}^{(2)} =-\frac{1}{2}\sum_{n=1}^{\infty}\sum_{\ell=0}^{2n}\sum_{k=0}^{\lfloor \frac{2n-\ell-1}{2}\rfloor}\sum_{q=0}^{\lfloor \frac{\ell-1}{2}\rfloor} &\frac{(-1)^{n+\ell}(2n+1)!\,\zeta_{2n+1}}{(2n-\ell-k)!\,(\ell-q)!\,k!\,q!}\left(\frac{\lambda}{16\pi^2}\right)^n  \notag \\
& \sqrt{2n-\ell-2k}\sqrt{\ell-2q}\, (\mathcal{P}_{2n-\ell-2k}^{+}\,\mathcal{P}_{\ell-2q}^{+}-\mathcal{P}_{2n-\ell-2k}^{-}\,\mathcal{P}_{\ell-2q}^{-}) \,.
\end{align}
Employing again the identity \eqref{integralidentity}, resumming the series in $\lambda$ in terms of the Bessel functions of the first kind and using the matrix elements defined in \eqref{Mnm}, we end up with
\begin{align}
\mathcal{M}_{\mathbb{Z}_2}^{(2)} = \frac{1}{2}\sum_{p,q=2}^{\infty}(-1)^{p-pq}\,\mathsf{M}_{p,q}\,(\mathcal{P}_{p}^{+}\,\mathcal{P}_{q}^{+}-\mathcal{P}_{p}^{-}\,\mathcal{P}_{q}^{-}) \,.
\end{align}

\subsection{The operator \texorpdfstring{$\mathcal{H}_{\mathbb{Z}_2}$}{}}

Using the change of basis \eqref{PP} and the \eqref{1ptA}, after few simple algebraic simplifications, we find
\begin{subequations}
\label{HZ2II} 
\begin{align}
& \mathcal{H}^{(0)}_{\mathbb{Z}_2} = -\frac{1}{12}\sum_{n=1}^{\infty}\sum_{\ell=0}^{n-1}(-1)^{n}\frac{\zeta_{2n+1}(2n+1)!}{(2n-2-2\ell)!\,(2\ell)!}\left(\frac{\lambda}{8\pi^2N}\right)^{n-1} t_{2\ell}\,t_{2n-2\ell-2}\,, \label{HZ2II0}\\
& \mathcal{H}^{(1)}_{\mathbb{Z}_2} = -\frac{\sqrt{2}}{12}\sum_{n=1}^{\infty}\sum_{\ell=0}^{n-1}\sum_{q=0}^{\ell}\sqrt{2\ell-2q}\,\frac{(-1)^{n}(2n+1)!\,\zeta_{2n+1}}{(2n-2-2\ell)!\,q!(2\ell-q)!}\left(\frac{\lambda}{8\pi^2N}\right)^{n-1} \left(\frac{N}{2}\right)^\ell  t_{2n-2\ell-2}\,\mathcal{P}^+_{2\ell-2q} \,, \label{HZ2II1} \\
& \mathcal{H}^{(2)}_{\mathbb{Z}_2} = - \frac{1}{24}\sum_{n=1}^{\infty}\sum_{\ell=0}^{2n-2}\sum_{k=0}^{\lfloor \frac{2n-\ell-3}{2}\rfloor}\sum_{q=0}^{\lfloor \frac{\ell-1}{2}\rfloor} \frac{(-1)^{n+\ell}(2n+1)!\,\zeta_{2n+1}}{(2n-2-\ell-k)!(\ell-q)!k!q!}\left(\frac{\lambda}{16\pi^2}\right)^{n-1}  \notag \\
& \qquad\qquad\qquad\qquad \sqrt{2n-\ell-2-2k}\sqrt{\ell-2q} \left(\mathcal{P}_{2n-2-\ell-2k}^{+}\,\mathcal{P}_{\ell-2q}^{+}-\mathcal{P}_{2n-2-\ell-2k}^{-}\,\mathcal{P}_{\ell-2q}^{-}\right) \label{HZ2II2} \,.
\end{align}
\end{subequations}
Inserting the \eqref{eventraces} into \eqref{HZ2II0} and using the identity \eqref{integralidentity}, after few straightforward steps we get
\begin{align}
\mathcal{H}^{(0)}_{\mathbb{Z}_2} = & \frac{4\pi^2}{3\lambda}N^2\int_0^\infty \frac{dt}{t}\,\frac{(t/2)^2}{\sinh(t/2)^2}\,\sum_{n=0}^{\infty}\sum_{\ell=0}^{n}\,\frac{(-1)^{n}}{(\ell+1)!\,\ell!\,(n-\ell)!\,(n-\ell+1)!}\left(\frac{\sqrt{\lambda}t}{4\pi}\right)^{2n+2} \notag \\
& -\frac{1}{12}\int_0^\infty dt\,t\,\frac{(t/2)^2}{\sinh(t/2)^2}\,\sum_{n=0}^{\infty}\sum_{\ell=0}^{n}\,\frac{(-1)^{n}}{\ell!\,(\ell-1)!\,(n-\ell)!\,(n-\ell+1)!}\left(\frac{\sqrt{\lambda}t}{4\pi}\right)^{2n} \notag \\
& +\frac{1}{72}\int_0^\infty dt\,t\,\frac{(t/2)^2}{\sinh(t/2)^2}\,\sum_{n=0}^{\infty}\sum_{\ell=0}^{n}\,\frac{(-1)^{n}}{\ell!\,(\ell-2)!\,(n-\ell)!\,(n-\ell+1)!}\left(\frac{\sqrt{\lambda}t}{4\pi}\right)^{2n} + O\left( N^{-2} \right)\,.
\end{align}
Then, suitably relabelling the summation indices and summing up the perturbative $\lambda$-series in terms of Bessel functions of the first kind \eqref{besselJ}, the expression for $\mathcal{H}_{\mathbb{Z}_2}^{(0)}$ can be rewritten in terms of the coefficients defined in \eqref{Mnm}-\eqref{Qnm} as
\begin{align}
\mathcal{H}_{\mathbb{Z}_2}^{(0)} = -\frac{4\pi^2}{3\lambda}\textsf{M}_{1,1}N^2 - \frac{1}{12}\left(\textsf{Q}_{1,1}-\frac{1}{6}\sum_{k=1}^{\infty}\sqrt{2k+1}\textsf{Q}_{1,2k+1}\right) + O(N^{-2}) \, ,
\end{align}
where we also used the identity \eqref{id:Q12} to rewrite the last term. \\
The same procedure can be applied also to $\mathcal{H}_{\mathbb{Z}_2}^{(1)}$, which, after plugging the leading large-$N$ contribution of \eqref{eventraces} into \eqref{HZ2II1}, becomes
\begin{align}
\mathcal{H}_{\mathbb{Z}_2}^{(1)} = \frac{\sqrt{2}}{12}N\int_0^\infty dt\,t\,\frac{(t/2)^2}{\sinh(t/2)^2}\,\sum_{n=0}^{\infty}\sum_{\ell=0}^{n}\sum_{q=0}^{\ell}\,\frac{(-1)^{n}\sqrt{2\ell-2q}\,\mathcal{P}^+_{2\ell-2q}}{q!\,(2\ell-q)!\,(n-\ell)!\,(n-\ell+1)!}\left(\frac{\sqrt{\lambda}t}{4\pi}\right)^{2n} +O\left(N^{-1} \right) \,,
\end{align}
and therefore, after using the expression \eqref{besselJ} and the matrix elements defined in \eqref{Q0n}, it comes out as
\begin{align}
\mathcal{H}_{\mathbb{Z}_2}^{(1)} = -\frac{\sqrt{2}N}{12}\sum_{k=1}^{\infty}\textsf{Q}_{0,2k}\,\mathcal{P}_{2k}^{+} + O(N^{-1}) \,.
\end{align}
Finally, utilising the identity \eqref{integralidentity} also in \eqref{HZ2II2} $\mathcal{H}_{\mathbb{Z}_2}^{(2)}$ becomes
\begin{align}
\mathcal{H}_{\mathbb{Z}_2}^{(2)} =\frac{1}{24}\int_0^\infty dt\,t\,\frac{(t/2)^2}{\sinh(t/2)^2}\, & \sum_{n=1}^{\infty}\sum_{\ell=0}^{2n-2} 
\sum_{k=0}^{\lfloor \frac{2n-\ell-1}{2}\rfloor}\sum_{q=0}^{\lfloor \frac{\ell-1}{2}\rfloor} \frac{(-1)^{n+\ell}}{(2n-\ell-k)!(\ell-q)!k!q!}\left(\frac{\sqrt{\lambda}t}{4\pi}\right)^{2n} \notag \\
& \sqrt{2n-\ell-2k}\sqrt{\ell-2q} (\mathcal{P}_{2n-\ell-2k}^{+}\,\mathcal{P}_{\ell-2q}^{+}-\mathcal{P}_{2n-\ell-2k}^{-}\,\mathcal{P}_{\ell-2q}^{-}) \,,
\end{align}
that, exactly as in the previous cases, can be resummed in terms of the Bessel functions yielding the following expression written using the coefficients \eqref{Qnm}
\begin{align}
\mathcal{H}_{\mathbb{Z}_2}^{(2)} = -\frac{1}{24}\sum_{p,q=2}^{\infty}(-1)^{p-pq}\,\textsf{Q}_{p,q}\,(\mathcal{P}^{+}_{p}\mathcal{P}^{+}_{q}-\mathcal{P}_{p}^{-}\mathcal{P}_{q}^{-}) \,.
\end{align}

\section{Identities and sum rules for \texorpdfstring{$\textsf{M}_{k,\ell}$}{} and \texorpdfstring{$\textsf{Q}_{k,\ell}$}{}}
\label{app:identities}

In this appendix, we prove the relations satisfied by the matrices $\textsf{M}_{k,\ell}$ and $\textsf{Q}_{k,\ell}$, which have been employed to determine the strong-coupling expansion of the integrated correlator $\mathcal{Y}$. Specifically, all these relations involve the resummation of an infinite sum of Bessel functions of the first kind. For this reason, we begin our analysis by considering the following identities, which will be used throughout this appendix
\begin{subequations}
\begin{align}
& J_{2n}(z) = \frac{2}{z} \sum_{k=n}^{\infty} (-1)^{n+k} (2k+1) J_{2k+1}(z) \,, \label{sumBesseleven} \\
& J_{2n+1}(z) = -\frac{2}{z} \sum_{k=n+1}^{\infty} (-1)^{n+k} (2k) J_{2k}(z) \label{sumbesselodd} \,, \\
& 2\sum_{k=1}^{\infty}(-1)^kk^3J_{2k}(z) = -\frac{z^2}{4}J_0(z)\, , \label{idk3} \\
& 2\sum_{k=1}^{\infty}(-1)^kk^5J_{2k}(z) = \frac{z^2}{4}\left(z\,J_1(z)-J_0(z)\right) \, . \label{idk5}
\end{align}
\label{identitiesBessel}
\end{subequations}
The identities \eqref{sumBesseleven} and \eqref{sumbesselodd} have already been exploited in Appendix A of \cite{Billo:2022fnb} and can be derived using the recurrence relation
\begin{align}
\label{recurrencerel}
J_{q-1}(z) = \frac{2\,q}{z} J_q(z) - J_{q+1}(z) \,.
\end{align}
On the other hand the identity \eqref{idk3} can be proven as follows
\begin{align}
& 2\sum_{k=1}^{\infty}(-1)^kk^3J_{2k}(z) = 2\sum_{k=1}^{\infty}(-1)^kk^3\sum_{m=0}^{\infty}\frac{(-1)^m}{m!\,(m+2k)!}\left(\frac{z}{2}\right)^{2m+2k} = \nonumber  \\[0.5em] 
& 2\sum_{n=1}^{\infty}(-1)^n\left(\frac{z}{2}\right)^{2n}\sum_{k=1}^{n}\frac{k^3}{(n-k)!\,(n+k)!} = \sum_{n=1}^{\infty}(-1)^n\left(\frac{z}{2}\right)^{2n}\frac{1}{(n-1)!\,(n-1)!}= \nonumber \\[0.5em]
& -\frac{z^2}{4}\sum_{m=0}^{\infty}(-1)^m\left(\frac{z}{2}\right)^{2m}\frac{1}{m!\,m!} = -\frac{z^2}{4}J_0(x)\, \ .
\end{align}
Finally the identity \eqref{idk5} can be proven in a completely similar manner.

\begin{itemize}
\item Proof of 
\begin{align}
\sum_{k=1}^{\infty}\sqrt{2k}\,\textsf{Q}_{0,2k} = -\textsf{Q}_{1,1}
\label{id:Q11}
\end{align}
Using the definition of $\textsf{Q}_{0,2k}$ given in \eqref{Q0n}, we obtain
\begin{align}
\sum_{k=1}^{\infty}\sqrt{2k}\,\textsf{Q}_{0,2k} & = - \sum_{k=1}^{\infty}(2k)(-1)^k \int_0^{\infty}\,dt\, t\, \frac{(t/2)^2}{\sinh(t/2)^2}\left(\frac{4\pi}{t\sqrt{\lambda}}\right)J_1\left(\frac{t\sqrt{\lambda}}{2\pi}\right)J_{2k}\left(\frac{t\sqrt{\lambda}}{2\pi}\right) \notag \\
& = \int_0^{\infty}\,dt\, t\, \frac{(t/2)^2}{\sinh(t/2)^2} J_1\left(\frac{t\sqrt{\lambda}}{2\pi}\right)J_{1}\left(\frac{t\sqrt{\lambda}}{2\pi}\right) = -\textsf{Q}_{1,1} \,,
\end{align}
where we have performed the sum over $k$ using the identity \eqref{sumbesselodd}.
\item Proof of
\begin{align}
\sum_{k=1}^{\infty}\sqrt{2k+1}\textsf{Q}_{1,2k+1} = \frac{1}{2}\int_0^{\infty}dt\, t\, \frac{(t/2)^2}{\sinh(t/2)^2}\left(\frac{t\sqrt{\lambda}}{2\pi}\right)J_1\left(\frac{t\sqrt{\lambda}}{2\pi}\right)J_2\left(\frac{t\sqrt{\lambda}}{2\pi}\right)
\label{id:Q12}
\end{align}
We use the definition of $\textsf{Q}_{1,2k+1}$ given in \eqref{Qnm} and then we perform the sums over $k$ by exploiting the identity \eqref{sumBesseleven}. In this manner, we obtain
\begin{align}
\sum_{k=1}^{\infty}\sqrt{2k+1}\textsf{Q}_{1,2k+1} & = -\sum_{k=1}^{\infty}(2k+1)(-1)^k \int_0^{\infty}\,dt\, t\, \frac{(t/2)^2}{\sinh(t/2)^2}\,J_1\left(\frac{t\sqrt{\lambda}}{2\pi}\right)J_{2k+1}\left(\frac{t\sqrt{\lambda}}{2\pi}\right) \notag \\
& = \frac{1}{2}\int_0^{\infty}dt\, t\, \frac{(t/2)^2}{\sinh(t/2)^2}\left(\frac{t\sqrt{\lambda}}{2\pi}\right)J_1\left(\frac{t\sqrt{\lambda}}{2\pi}\right)J_2\left(\frac{t\sqrt{\lambda}}{2\pi}\right) \,.
\end{align}
\item Proof of 
\begin{align}
& \sum_{k=1}^{\infty}\sqrt{k}\,k^2\,\textsf{M}_{0,2k} = -\frac{1}{2\sqrt{2}}(1+\lambda\partial_{\lambda})\textsf{M}_{1,1}
\label{idM11FIRST}
\end{align}
Using \eqref{M0n} we rewrite the left-hand side as
\begin{align}
\sum_{k=1}^{\infty}\sqrt{k}\,k^2\,\textsf{M}_{0,2k} & = -\sqrt{2}\int_0^{\infty}\frac{dt}{t}\,\frac{(t/2)^2}{\sinh(t/2)^2}\left(\frac{4\pi}{t\sqrt{\lambda}}\right)\,J_1\left(\frac{t\sqrt{\lambda}}{2\pi}\right)\sum_{k=1}^{\infty}(-1)^kk^3J_{2k}\left(\frac{t\sqrt{\lambda}}{2\pi}\right)  \nonumber \\
& = \frac{1}{\sqrt{2}} \int_0^{\infty} \frac{dt}{t} \frac{(t/2)^2}{\sinh(t/2)^2}\, J_1\left(\frac{t\sqrt{\lambda}}{2\pi}\right) \left(\frac{t\sqrt{\lambda}}{4\pi}\right)J_0\left(\frac{t\sqrt{\lambda}}{2\pi}\right)\,  , 
\label{M11step1}
\end{align}
where the sum over $k$ has been performed using the identity \eqref{idk3}.
Then we use the following relation
\begin{align}
\left(\frac{t\sqrt{\lambda}}{4\pi}\right)J_0\left(\frac{t\sqrt{\lambda}}{2\pi}\right) = \frac{1}{2}J_1\left(\frac{t\sqrt{\lambda}}{2\pi}\right) + \lambda\partial_{\lambda}J_1\left(\frac{t\sqrt{\lambda}}{2\pi}\right)\, . 
\end{align}
This manner the expression \eqref{M11step1} becomes 
\begin{align}
\frac{1}{2\sqrt{2}} \int_0^{\infty} \frac{dt}{t} \frac{(t/2)^2}{\sinh(t/2)^2}\, J_1\left(\frac{t\sqrt{\lambda}}{2\pi}\right)\left(1 + 2\lambda\partial_{\lambda}\right)J_1\left(\frac{t\sqrt{\lambda}}{2\pi}\right) = -\frac{1}{2\sqrt{2}}(1+\lambda\partial_{\lambda})\textsf{M}_{1,1}\, ,   
\end{align}
where we used \eqref{Mnm}.
\item Proof of
\begin{align}
\sum_{k=1}^{\infty}\sum_{\ell=1}^{\infty}\sqrt{k\ell}\,\textsf{M}_{0,2k}\,\textsf{M}_{0,2\ell}\,(k^2+\ell^2) = \frac{1}{4}(2+\lambda\partial_{\lambda})\textsf{M}_{1,1}^2
\label{idM11SECOND}
\end{align}
we rewrite the left-hand side as
\begin{align} 2\sum_{k=1}^{\infty}\sqrt{k}\,\textsf{M}_{0,2k}\sum_{\ell=1}^{\infty}\sqrt{\ell}\,\ell^2\textsf{M}_{0,2\ell}\nonumber = \frac{\textsf{M}_{1,1}^2}{2}+ \frac{\textsf{M}_{1,1}}{2}\lambda\partial_{\lambda}\textsf{M}_{1,1} = \frac{1}{4}(2+\lambda\partial_{\lambda})\textsf{M}_{1,1}^2\,  ,    
\end{align}
where the summations over $\ell$ and $k$ have been performed using the identity \eqref{idM11FIRST} and the identity \eqref{idM11}, respectively.
\item Proof of
\begin{align}
& \sum_{k=1}^{\infty}\sum_{\ell=1}^{\infty}\sqrt{k\,\ell}(k^4+\ell^4)\textsf{M}_{0,2k}\textsf{M}_{0,2\ell} = \nonumber \\
& - \textsf{M}_{1,1}\frac{\sqrt{\lambda}}{8\pi^2} \int_0^{\infty}dt\, \frac{(t/2)^2}{\sinh(t/2)^2}J_1\left(\frac{t\sqrt{\lambda}}{2\pi}\right)\left[2\pi J_0\left(\frac{t\sqrt{\lambda}}{2\pi}\right)-t\sqrt{\lambda}J_1\left(\frac{t\sqrt{\lambda}}{2\pi}\right)\right]\, .
\label{idM11THIRD}
\end{align}
We rewrite the left-hand side as
\begin{align}
2\sum_{\ell=1}^{\infty}\sqrt{\ell}\,\textsf{M}_{0,2\ell}\sum_{k=1}^{\infty}\sqrt{k}\,k^4\,\textsf{M}_{0,2k} = -\textsf{M}_{1,1}\sum_{k=1}^{\infty}\sqrt{2k}\,k^4\,\textsf{M}_{0,2k}\,  ,  
\end{align}
where the sum over $\ell$ has been performed using \eqref{idM11}. Then, we use \eqref{M0n} and we obtain
\begin{align}
& -\textsf{M}_{1,1}\sum_{k=1}^{\infty}\sqrt{2k}\,k^4\,\textsf{M}_{0,2k} = \textsf{M}_{1,1}\int_0^{\infty}\frac{dt}{t}\frac{(t/2)^2}{\sinh(t/2)^2}\,J_1\left(\frac{t\sqrt{\lambda}}{2\pi}\right)\left(\frac{8\pi}{t\sqrt{\lambda}}\right)\sum_{k=1}^{\infty}(-1)^k\,k^5J_{2k}\left(\frac{t\sqrt{\lambda}}{2\pi}\right) \nonumber \\[0.5em]
&  = - \textsf{M}_{1,1}\frac{\sqrt{\lambda}}{8\pi^2} \int_0^{\infty}dt\, \frac{(t/2)^2}{\sinh(t/2)^2}\,J_1\left(\frac{t\sqrt{\lambda}}{2\pi}\right)\left[2\pi J_0\left(\frac{t\sqrt{\lambda}}{2\pi}\right)-t\sqrt{\lambda}J_1\left(\frac{t\sqrt{\lambda}}{2\pi}\right)\right] \, ,
\end{align}
where the sum over $k$ has been performed using the identity \eqref{idk5}.
\end{itemize}

\section{Analytic evaluations at strong-coupling}
\label{app:StrongCoupling}
In this appendix, we outline the method employed to derive certain strong-coupling expansions discussed in Section \ref{subsec:JJJJNLO}. As a first step we introduce the new variables
\begin{align}
t = \frac{\sqrt{x}}{2g}\, \quad \qquad \text{with} \quad \qquad g=\frac{\sqrt{\lambda}}{4\pi}\,  .    
\end{align}
Then, the matrices \eqref{Xmatrix}, \eqref{Mnm} and \eqref{Qnm} can be rewritten as follows
\begin{subequations}
\begin{align}
& \textsf{X}_{n,m} = (-1)^{\frac{n+m+2nm}{2}}\sqrt{n\,m}\int_0^{\infty}\frac{dx}{x}\,\chi\left(\frac{\sqrt{x}}{2g}\right)J_n(\sqrt{x})\,J_{m}(\sqrt{x})\, ,\label{Xnew} \\[0.5em]
& \textsf{M}_{n,m} = (-1)^{\frac{n+m+2nm}{2}}\frac{\sqrt{nm}}{32g^2}\int_0^{\infty}dx\, \chi\left(\frac{\sqrt{x}}{2g}\right)J_n(\sqrt{x})\,J_{m}(\sqrt{x})\, , \label{Mnew} \\[0.5em]
& \textsf{Q}_{n,m} = (-1)^{\frac{n+m+2nm}{2}}\frac{\sqrt{n\,m}}{128g^4}\int_0^{\infty}dx\, x\, \chi\left(\frac{\sqrt{x}}{2g}\right)J_n(\sqrt{x})\,J_m(\sqrt{x})\, , \label{Qnew}
\end{align}
\end{subequations}
where 
\begin{align}
\chi(x) = -\frac{1}{\sinh(x/2)^2}\,  .    
\end{align}

\subsection{Strong-coupling behaviour of \texorpdfstring{$-\text{Tr}[\text{Q}(\mathbb{1}-\text{D})]$}{}}
\label{app:StrongCouplingQD}
Here we derive the leading strong-coupling behaviour of
\begin{align}
& -\sum_{p,q=2}^{\infty}(-1)^{p-pq}\textsf{Q}_{p,q}(\delta_{p,q}-\textsf{D}_{p,q}) = -\text{Tr}[\textsf{Q}(\mathbb{1}-\textsf{D})] = \sum_{k=1}^{\infty}\text{Tr}[\textsf{Q}\,(\textsf{X})^k] \nonumber \\
& = \sum_{k=1}^{\infty}\text{Tr}[\textsf{Q}^{\text{even}}(\textsf{X}^{\text{even}})^k]
+ \sum_{k=1}^{\infty}\text{Tr}[\textsf{Q}^{\text{odd}}(\textsf{X}^{\text{odd}})^k]
\, , \label{Tr(Q1D)} 
\end{align}
where we used the expressions \eqref{Xall} and \eqref{Qall}.
Let us start by considering the case $k=1$. We employ both \eqref{Qnew} and \eqref{Xnew} and we obtain
\begin{subequations}
\begin{align}
\text{Tr}[\textsf{Q}^{\text{even}}\,\textsf{X}^{\text{even}}] & = \frac{1}{128g^4}\int_0^{\infty}dx\,x \int_0^{\infty}\frac{dy}{y}\,\chi\left(\frac{\sqrt{x}}{2g}\right)\chi\left(\frac{\sqrt{y}}{2g}\right)\left[\sum_{p=1}^{\infty}(2p)\,J_{2p}(\sqrt{x})J_{2p}(\sqrt{y})\right]^2  \nonumber \\  
& = \frac{1}{128g^4}\int_0^{\infty}dx\, x^2 \int_0^{\infty} dy\,\chi\left(\frac{\sqrt{x}}{2g}\right)\chi\left(\frac{\sqrt{y}}{2g}\right)K^{\text{even}}(x,y)^2 \, \ , \\[0.5em]
\text{Tr}[\textsf{Q}^{\text{odd}}\,\textsf{X}^{\text{odd}}] & = \frac{1}{128g^4}\int_0^{\infty}dx\,x \int_0^{\infty}\frac{dy}{y}\,\chi\left(\frac{\sqrt{x}}{2g}\right)\chi\left(\frac{\sqrt{y}}{2g}\right)\left[\sum_{p=1}^{\infty}(2p+1)\,J_{2p+1}(\sqrt{x})J_{2p+1}(\sqrt{y})\right]^2  \nonumber \\  
& = \frac{1}{128g^4}\int_0^{\infty}dx\, x^2 \int_0^{\infty} dy\,\chi\left(\frac{\sqrt{x}}{2g}\right)\chi\left(\frac{\sqrt{y}}{2g}\right)K^{\text{odd}}(x,y)^2 \, ,
\end{align}
\label{TrQX}
\end{subequations}
where we introduced the Bessel kernels
\begin{subequations}
\begin{align}
& K^{\text{odd}}(x,y) = \frac{1}{\sqrt{xy}}\sum_{p=1}^{\infty}(2p+1)J_{2p+1}(\sqrt{x})J_{2p+1}(\sqrt{y}) = \frac{\sqrt{x}J_3(\sqrt{x})J_2(\sqrt{y})-\sqrt{y}J_3(\sqrt{y})J_2(\sqrt{x})}{2(x-y)}\, , \label{Kodd} \\
& K^{\text{even}}(x,y) = \frac{1}{\sqrt{xy}}\sum_{p=1}^{\infty}(2p)J_{2p}(\sqrt{x})J_{2p}(\sqrt{y}) =  \frac{\sqrt{x}J_2(\sqrt{x})J_1(\sqrt{y})-\sqrt{y}J_2(\sqrt{y})J_1(\sqrt{x})}{2(x-y)}\, . \label{Keven}
\end{align}
\label{Kall}
\end{subequations}
As discussed in \cite{Belitsky:2020qrm,Belitsky:2020qir}, at the leading order of the strong-coupling expansion, we can approximate the expressions \eqref{Kall} by considering their asymptotic behaviours for large values of their arguments, namely
\begin{subequations}
\begin{align}
& K^{\text{odd}}(x,y) \, \simeq \, \frac{1}{2\pi(xy)^{1/4}}\left[\frac{\sin(\sqrt{x}-\sqrt{y})}{\sqrt{x}-\sqrt{y}}-\frac{\cos(\sqrt{x}+\sqrt{y})}{\sqrt{x}+\sqrt{y}}\right]\, , \\[0.5em]
& K^{\text{even}}(x,y) \, \simeq \, \frac{1}{2\pi(xy)^{1/4}}\left[\frac{\sin(\sqrt{x}-\sqrt{y})}{\sqrt{x}-\sqrt{y}}+\frac{\cos(\sqrt{x}+\sqrt{y})}{\sqrt{x}+\sqrt{y}}\right]\,  .
\end{align}
\label{KallStrong}
\end{subequations}
We observe that the two expressions above differ only by the sign in front of the cosine term. Importantly, this term oscillates rapidly over the integration domain and, hence, it can be neglected since it does not contribute to the integral. Therefore, the expressions \eqref{KallStrong} are equivalent at strong-coupling. Furthermore, following the procedure outlined in \cite{Belitsky:2020qrm,Belitsky:2020qir}, one can demonstrate that at the leading order of the strong-coupling expansion, for any test function $f$, the following relation holds
\begin{align}
\int_0^{\infty}dy\, K^{\text{odd}}(x,y)\,\chi\left(\frac{\sqrt{y}}{2g}\right)\,f(y) \simeq \int_0^{\infty}dy\, K^{\text{even}}(x,y)\,\chi\left(\frac{\sqrt{y}}{2g}\right)\,f(y) \simeq \chi\left(\frac{\sqrt{x}}{2g}\right)f(x) \, .
\label{IdentityDelta}
\end{align}
Hence, applying \eqref{IdentityDelta} to the integrals \eqref{TrQX} results in
\begin{subequations}
\begin{align}
\text{Tr}[\textsf{Q}^{\text{even}}\,\textsf{X}^{\text{even}}] & \simeq \frac{1}{128g^4}\int_0^{\infty}dx\,x^2\,\chi\left(\frac{\sqrt{x}}{2g}\right)^2K^{\text{even}}(x,x)\, ,   \\[0.5em]
\text{Tr}[\textsf{Q}^{\text{odd}}\,\textsf{X}^{\text{odd}}] & \simeq \frac{1}{128g^4}\int_0^{\infty}dx\,x^2\,\chi\left(\frac{\sqrt{x}}{2g}\right)^2K^{\text{odd}}(x,x)\, .   
\end{align}
\end{subequations}
Since the above relations demonstrate equivalence between the even and the odd case at strong-coupling, from now on, for the sake of simplicity, we will focus solely on the even case.
We can then replicate the same steps employed for the $k=1$ case for any arbitrary value of $k$, thereby obtaining
\begin{align}
\text{Tr}[\textsf{Q}^{\text{even}}\,(\textsf{X}^{\text{even}})^k] \simeq   \frac{1}{128g^4}\int_0^{\infty}dx\, x^2\, \chi\left(\frac{\sqrt{x}}{2g}\right)^{1+k}K^{\text{even}}(x,x)\, .      
\end{align}
Then, the strong-coupling limit of the even contribution in the expression \eqref{Tr(Q1D)} is 
\begin{align}
\sum_{k=1}^{\infty}\text{Tr}[\textsf{Q}^{\text{even}}\,(\textsf{X}^{\text{even}})^k] \simeq \frac{1}{128\,g^4}\int_0^{\infty}dx\, x^2\, \frac{\chi\left(\frac{\sqrt{x}}{2g}\right)^2}{1-\chi\left(\frac{\sqrt{x}}{2g}\right)}\,K^{\text{even}}(x,x)\, .    
\end{align}
As a final step, we perform the change of variables $x=4g^2z^2$. Furthermore, from \eqref{KallStrong}, it follows that
\begin{align}
K^{\text{odd}}(x,x) \, \simeq \, K^{\text{even}}(x,x) \, \simeq \, \frac{1}{2\pi\sqrt{x}}\, .
\label{KStrong}
\end{align}
This manner, we finally obtain
\begin{align}
-\text{Tr}[\textsf{Q}^{\text{even}}(\mathbb{1}-\textsf{D}^{\text{even}})] \, \simeq \, \frac{\sqrt{\lambda}}{16\pi^2}\int_0^{\infty} dz\, z^4\, \frac{\chi(z)^2}{1-\chi(z)} \, \underset{\lambda \rightarrow \infty}{\sim} \, \frac{\pi^2\sqrt{\lambda}}{120} + O(\lambda^0)\, .   
\end{align}
Therefore we conclude that the leading term of the strong-coupling expansion of \eqref{Tr(Q1D)} reads
\begin{align}
-\text{Tr}[\textsf{Q}^{\text{even}}(\mathbb{1}-\textsf{D}^{\text{even}})]-\text{Tr}[\textsf{Q}^{\text{odd}}(\mathbb{1}-\textsf{D}^{\text{odd}})]  \, \underset{\lambda \rightarrow \infty}{\sim} \, \frac{\pi^2\sqrt{\lambda}}{60} + O(\lambda^0)\, .    
\end{align}

\subsection{Strong-coupling behaviour of \texorpdfstring{$\text{Tr}[(\text{M}^{\text{even}})^2]$}{} and \texorpdfstring{$\text{Tr}[(\text{M}^{\text{odd}})^2]$}{}}
\label{app:TrM2strong}
Here we derive the leading strong-coupling behaviour of $\text{Tr}[(\textsf{M}^{\text{even}})^2]$ and $\text{Tr}[(\textsf{M}^{\text{odd}})^2]$.
We use \eqref{Mnew} to obtain
\begin{align}
\text{Tr}[(\textsf{M}^{\text{even}})^2]& = \frac{1}{(32g^2)^2}\int_0^{\infty}dx \int_0^{\infty} dy\, \chi\left(\frac{\sqrt{x}}{2g}\right)\chi\left(\frac{\sqrt{y}}{2g}\right)\left[\sum_{\ell=1}^{\infty}2\ell\,J_{2\ell}(\sqrt{x})J_{2\ell}(\sqrt{y})\right]^2 \nonumber \\
& = \frac{1}{(32g^2)^2}\int_0^{\infty}dx\, x \int_0^{\infty}dy\, y\, \chi\left(\frac{\sqrt{x}}{2g}\right)\chi\left(\frac{\sqrt{y}}{2g}\right) \left[K^{\text{even}}(x,y)\right]^2 \,.
\label{TrM2step1}
\end{align}
Then, we employ the relation \eqref{IdentityDelta}, we perform the change of variables $x=4g^2z^2$ and we approximate $K^{\text{even}}(x,x)$ as \eqref{KStrong}. In this way we obtain
\begin{align}
\text{Tr}[(\textsf{M}^{\text{even}})^2]\,\simeq \,  \frac{g}{32\pi}\int_0^{\infty}dz\, z^4 \, \chi(z)^2 \, \underset{\lambda \rightarrow \infty}{\sim} \, \sqrt{\lambda}\left(\frac{1}{12}-\frac{\pi ^2}{180}\right) + O(\lambda^0)\, .
\label{TrM2evenStrong}
\end{align}
The computation of $\text{Tr}[(\textsf{M}^{\text{odd}})^2]$ can be performed following the same steps outlined above. Moreover, considering the relations \eqref{IdentityDelta} and \eqref{KStrong}, the resulting expression coincides with \eqref{TrM2evenStrong}.
Hence, at the leading order of the strong-coupling expansion, it holds that
\begin{align}
\,\left(\text{Tr}[(\textsf{M}^{\text{even}})^2]+\text{Tr}[(\textsf{M}^{\text{odd}})^2]\right) \, \underset{\lambda \rightarrow \infty}{\sim} \, \frac{\sqrt{\lambda}}{2}\left(\frac{1}{3}-\frac{\pi ^2}{45}\right) + O(\lambda^{0})\,  .
\label{resultI1}
\end{align}

%\subsection{Strong coupling behaviour of $\mathcal{I}^{(1)}$}
%\label{app:I1}
%Here we derive the strong coupling behaviour of $\mathcal{I}^{(1)}$ as defined in \eqref{con1}. We use the expression \eqref{Mnew} and we obtain
%\begin{align}
%\mathcal{I}^{(1)} & =  \frac{1}{(32g^2)^2}\int_0^{\infty}dx\,\int_0^{\infty}dy\, \chi\left(\frac{\sqrt{x}}{2g}\right)\,  \chi\left(\frac{\sqrt{y}}{2g}\right)\left[\sum_{q=1}^{\infty}q\,J_q(\sqrt{x})J_{q}(\sqrt{y})\right]^2 = \nonumber \\[0.5em]
%& = \frac{4}{(32g^2)}\int_0^{\infty}dx\, x^2\, \chi\left(\frac{\sqrt{x}}{2g}\right)^2K^{\text{even}}(x,x)\, \ ,
%\end{align}   
%where we employed the expression \eqref{Kkernel} and we performed the integral over $y$ using the relation \eqref{IdentityDelta}. We observe that the expression above coincides, up to a numerical factor, with \eqref{TrM2step1}. Therefore, the leading term of its strong coupling expansion reads
%\begin{align}
%\mathcal{I}^{(1)} \underset{\lambda \rightarrow \infty}{\sim} \sqrt{\lambda}\left(\frac{1}{3}-\frac{\pi ^2}{45}\right) + O(\lambda^{0})\, \ . 
%\end{align}

\subsection{Strong-coupling behaviour of \texorpdfstring{$\mathcal{I}^{(2)}$}{}}
\label{app:I2}
Here, we derive the leading term of the strong-coupling expansion of $\mathcal{I}^{(2)}$, as defined in \eqref{I2andI3}. It is evident that by employing \eqref{Xall}, the expression \eqref{con2} for $k_1=1$ can be rewritten as
\begin{align}
\mathcal{I}_1^{(2)} = \sum_{q_1,p_1=1}^{\infty}\sum_{q_2=1}^{\infty}\,\left[\textsf{M}^{\text{even}}_{p_1,q_1}\,\textsf{M}^{\text{even}}_{q_1,q_2}\,\textsf{X}^{\text{even}}_{q_2,p_1} + \textsf{M}^{\text{odd}}_{p_1,q_1}\,\textsf{M}^{\text{odd}}_{q_1,q_2}\,\textsf{X}^{\text{odd}}_{q_2,p_1}\,\right] .
\end{align}
For the time being let us restrict our attention to the contribution involving $\textsf{X}^{\text{even}}$. We use \eqref{Xnew} and \eqref{Mnew} and we obtain
\begin{align}
& \frac{1}{(32g^2)^2} \int_0^{\infty}dx_1\int_0^{\infty}dx_2\int_0^{\infty}\frac{dy}{y}\,\chi\left(\frac{\sqrt{x_1}}{2g}\right)\,  \chi\left(\frac{\sqrt{x_2}}{2g}\right)\,  \chi\left(\frac{\sqrt{y}}{2g}\right)\left[\sum_{p_1=1}^{\infty}2p_1J_{2p_1}(\sqrt{x_1})J_{2p_1}(\sqrt{x_2})\right]\times\nonumber \\
& \left[\sum_{q_1=1}^{\infty}2q_1J_{2q_1}(\sqrt{x_1})J_{2q_1}(\sqrt{y})\right]\left[\sum_{q_2=1}^{\infty}2q_2J_{2q_2}(\sqrt{y})J_{2q_2}(\sqrt{x_2})\right]  = \frac{1}{(32g^2)^2}\int_0^{\infty}dx_1\, x_1^2\, \chi\left(\frac{\sqrt{x_1}}{2g}\right)^3K^{\text{even}}(x_1,x_1)
\label{I2step1}
\end{align}
where we used \eqref{Kall} and we performed the integration over $x_2$ and $y$ using the relation \eqref{IdentityDelta}. It is straightforward to observe that if we had instead considered the contribution involving the $\textsf{X}^{\text{odd}}$ matrix, the only difference would have been the replacement of $K^{\text{even}}(x_1,x_1)$ with $K^{\text{odd}}(x_1,x_1)$ in \eqref{I2step1}. Furthermore, the case with arbitrary $k_1$ can be analysed in a similar manner, and it reads
\begin{align}
\mathcal{I}_{k_1}^{(2)} =  \frac{1}{(32g^2)^2}\int_0^{\infty}dx\, x^2\, \chi\left(\frac{\sqrt{x}}{2g}\right)^{2+k_1}\left[K^{\text{even}}(x,x)+K^{\text{odd}}(x,x)\right]\, \ .      
\end{align}
Therefore, the strong coupling expansion of $\mathcal{I}^{(2)}$ can be obtained by considering
\begin{align}
\sum_{k_1=1}^{\infty}\mathcal{I}_{k_1}^{(2)} = \frac{1}{(32g^2)^2}\int_0^{\infty}dx\, x^2 \frac{\chi\left(\frac{\sqrt{x}}{2g}\right)^3}{1-\chi\left(\frac{\sqrt{x}}{2g}\right)}\,[K^{\text{even}}(x,x)+K^{\text{odd}}(x,x)]\,  .     
\end{align}
At this point we perform the change of variables $x=4g^2z^2$ and we use \eqref{KStrong}. In this manner, we finally obtain
\begin{align}
\mathcal{I}^{(2)} \, \simeq \, \frac{\sqrt{\lambda}}{64\,\pi^2}\int_0^{\infty}dz\, z^4 \frac{\chi(z)^3}{1-\chi(z)}\, \underset{\lambda \rightarrow \infty}{\sim} \,\frac{\sqrt{\lambda}}{2}\left(\frac{19 \pi ^2}{720}-\frac{1}{3}\right)  +O(\lambda^{0})\, . 
\label{partial1}
\end{align}

\subsection{Strong-coupling behaviour of \texorpdfstring{$\mathcal{I}^{(3)}$}{}}
\label{app:I3}
Here, we compute the leading term of the strong-coupling expansion of $\mathcal{I}^{(3)}$, as defined in \eqref{I2andI3}. We use \eqref{Xall}
and we rewrite the expression \eqref{con3} with $k_1=k_2=1$ as
\begin{align}
\mathcal{I}_{1,1}^{(3)}  = \sum_{q_1,p_1=1}^{\infty}\sum_{q_2,p_2=1}^{\infty}\,\left[\textsf{M}^{\text{even}}_{p_1,q_1}\,\textsf{X}^{\text{even}}_{q_1,p_2}\,\textsf{M}^{\text{even}}_{p_2,q_2}\,\textsf{X}^{\text{even}}_{q_2,p_1} + \textsf{M}^{\text{odd}}_{p_1,q_1}\,\textsf{X}^{\text{odd}}_{q_1,p_2}\,\textsf{M}^{\text{odd}}_{p_2,q_2}\,\textsf{X}^{\text{odd}}_{q_2,p_1}\right]\,  .
\label{I3start}
\end{align}
Let us begin by considering  the contribution in \eqref{I3start} involving the insertion of two $\textsf{X}^{\text{even}}$ matrices. We employ the same procedure used for the evaluation of \eqref{I2step1} and we obtain
\begin{align}
\sum_{q_1,p_1=1}^{\infty}\sum_{q_2,p_2=1}^{\infty} \textsf{M}^{\text{even}}_{p_1,q_1}\,\textsf{X}^{\text{even}}_{q_1,p_2}\,\textsf{M}^{\text{even}}_{p_2,q_2}\,\textsf{X}^{\text{even}}_{q_2,p_1} \simeq \frac{1}{(32g^2)^2}\int_{0}^{\infty}dx\, x^2\, \chi\left(\frac{\sqrt{x}}{2g}\right)^4K^{\text{even}}(x,x)\, .
\label{I3step1}
\end{align}
The contribution in \eqref{I3start} involving the insertion of two $\textsf{X}^{\text{odd}}$ matrices can be worked out in a similar way. Therefore we find
\begin{align}
\mathcal{I}_{1,1}^{(3)} = \frac{1}{(32g^2)^2}\int_{0}^{\infty}dx\, x^2\, \chi\left(\frac{\sqrt{x}}{2g}\right)^4[K^{\text{even}}(x,x)+K^{\text{odd}}(x,x)]\, .  
\label{I3step2}
\end{align}
The case with arbitrary $k_1$ and $k_2$ can be easily worked out using the same procedure used to derive \eqref{I3step2}, yielding
\begin{align}
\mathcal{I}_{k_1,k_2}^{(3)} = \frac{1}{(32g^2)^2}\int_{0}^{\infty}dx\, x^2\, \chi\left(\frac{\sqrt{x}}{2g}\right)^{k_1+k_2+2}[K^{\text{even}}(x,x)+K^{\text{odd}}(x,x)]\, .
\end{align}
Then, the sums over $k_1$ and $k_2$ can be done as follows %\textcolor{blue}{credo non ci siano le sommatotorie su $k_1$ e $k_2$} \textcolor{red}{Hai ragione mi ero scordato di scrivere $\mathcal{I}_{k_1,k_2}^{(3)}$, se ritieni che sia troppo pedante scrivere anche questo passaggio possiamo toglierlo :-)}
\begin{align}
\mathcal{I}^{(3)} = \sum_{k_1=1}^{\infty}\sum_{k_2=1}^{\infty} \mathcal{I}_{k_1,k_2}^{(3)}=  \frac{1}{(32g^2)^2}\int_0^{\infty}dx\, x^2 \frac{\chi\left(\frac{\sqrt{x}}{2g}\right)^4}{\left(1-\chi\left(\frac{\sqrt{x}}{2g}\right)\right)^2}\,[K^{\text{even}}(x,x) +K^{\text{odd}}(x,x)]\, .
\end{align}
We finally perform the change of variables $x=4g^2z^2$ and we use the relation \eqref{KStrong}. This way we get our final expression  
\begin{align}
\mathcal{I}^{(3)} \, \simeq \, \frac{\sqrt{\lambda}}{64\pi^2}\int_0^{\infty}dz\, z^4 \frac{\chi(z)^4}{(1-\chi(z))^2} \, \underset{\lambda \rightarrow \infty}{\sim} \, \frac{\sqrt{\lambda}}{2}\left(\frac{1}{6}-\frac{\pi^2}{90}\right) + O(\lambda^0)\, .    
\end{align}

\section{Numerical evaluation of the expression \texorpdfstring{\eqref{M2M1c2}}{}}
\label{app:numerics}
\begin{figure}
    \centering
    \includegraphics[scale=0.4]{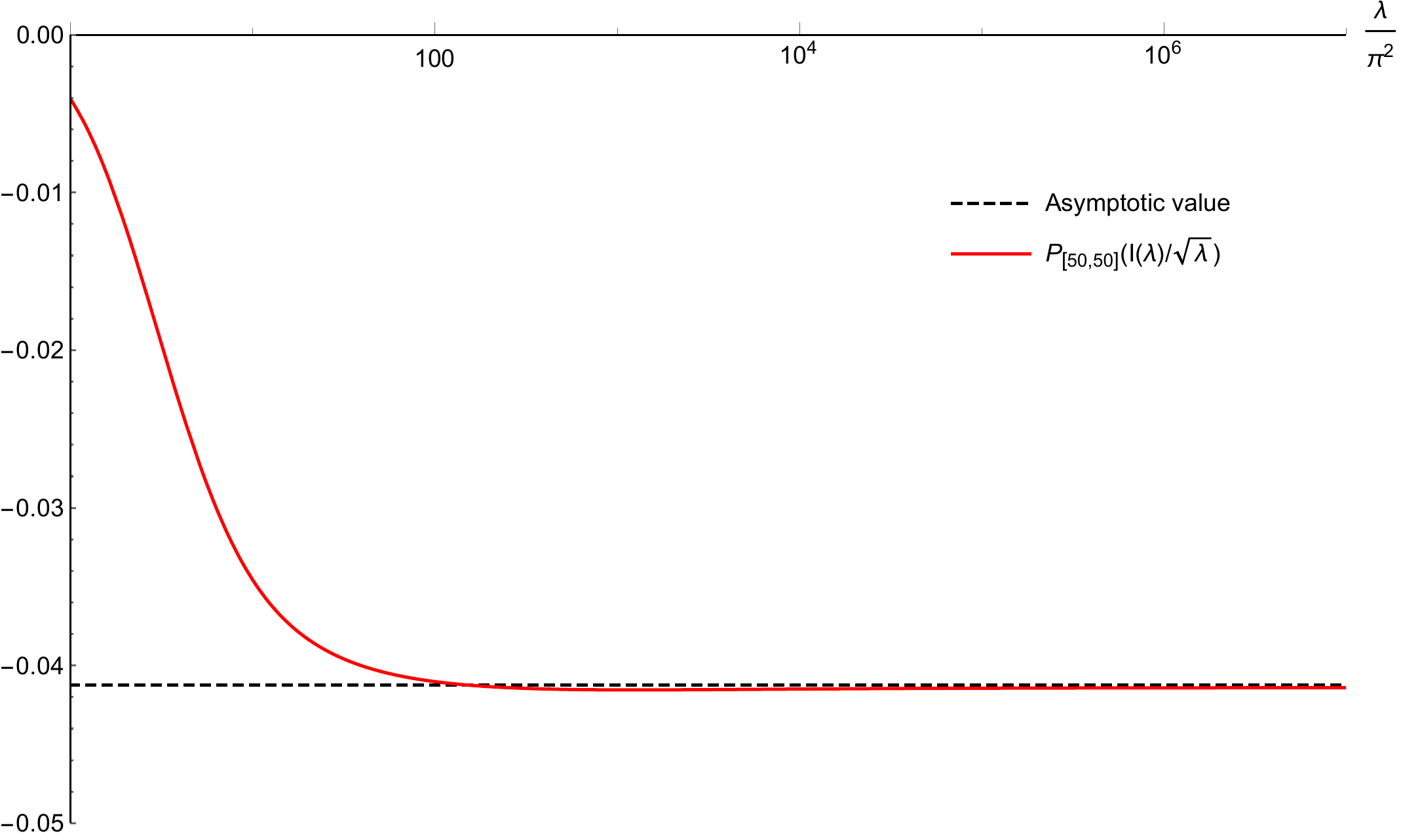}
    \caption{The red curve represents a diagonal Padé approximant of degree 50 for the perturbative expansion of $\mathcal{I}(\lambda)/\sqrt{\lambda}$. The dashed black line represents its asymptotic constant value, which, estimated within numerical errors, is compatible with $-\pi^2/240$.}
    \label{fig:Pade}
\end{figure}

In this appendix, we elaborate on our method for numerically evaluating the expression \eqref{M2M1c2} for large values of the 't Hooft coupling $\lambda$. Let us begin by introducing the quantity
\begin{align}
\mathcal{I}(\lambda) \equiv \sum_{p,q=2}^{\infty}(-1)^{p-pq}\textsf{M}_{p,q}(\sqrt{p\,q}-\textsf{d}_p\textsf{d}_q) \, .   
\label{II}
\end{align}
In the perturbative regime (i.e. $\lambda \ll 1$) by employing the expression \eqref{Mnm}, along with the definition \eqref{dandD} of the $\textsf{d}_{p}$ coefficients, we generated a very long series expansion of \eqref{II} and we reached the order $O(\lambda^{100})$. For example, the first few terms of such  expansion are
\begin{align}
\mathcal{I}(\lambda)   = -\frac{45\, \zeta_3\, \zeta_5}{512 \pi ^8}\,\lambda^4 + \frac{75 \left(10\, \zeta_5^2+21\, \zeta_3\, \zeta_7\right)}{8192 \pi
   ^{10}}\lambda^5  + O(\lambda^5)\,  .
\end{align}
Then, starting from the aforementioned perturbative series, we can construct a Padé approximant that enables us to extend the series beyond its radius of convergence, which is located at $\lambda \simeq \pi^2$, and extract information in the strong-coupling regime. A particular choice, which has proven to be useful for various observables (see, for example \cite{Pini:2023svd,Beccaria:2021vuc}), is provided by a diagonal conformal Padé approximant. This means that before computing the Padé approximant, we perform a conformal map, namely
\begin{align}
    \frac{\lambda}{\pi^2} \mapsto \frac{4z}{(z-1)^2}\, .
\end{align}
Then, we construct a Padé approximant in the new variable $z$ inside the unit circle $|z| \leq 1$, and finally, we revert to the initial variable $\lambda$ using the map 
\begin{align}
z = \frac{\sqrt{1+\frac{\lambda}{\pi^2}}-1}{\sqrt{1+\frac{\lambda}{\pi^2}}+1}\,  .    
\end{align}
In this manner, as discussed in \cite{Costin:2019xql,Costin:2020hwg}, we obtain a Padé approximant that remains stable for very large values of the 't Hooft coupling.
The outcome of this procedure is reported in Figure \ref{fig:Pade}. Based on this result, we are consequently led to provide the following numerical estimation for \eqref{II}
\begin{align}
\mathcal{I}(\lambda) \, \underset{\lambda \rightarrow \infty}{\sim} \,  -\frac{\pi^2}{240}\sqrt{\lambda} + O(\lambda^0) \, .
\end{align}

\section{Derivation of the relation \texorpdfstring{\eqref{Tplus}}{}}
\label{app:Tplus}
In this appendix, we argue for the validity of the relation \eqref{Tplus} and provide a formal proof for the terms linear in $\zeta$ in the perturbative expansion. As an initial step, we use \eqref{Pn} to obtain
\begin{align}
& \langle \mathcal{P}_{2k_1}^{+}\mathcal{P}_{2k_2}^{+} \rangle = 2\sqrt{k_1k_2}\left(\frac{2}{N}\right)^{k_1+k_2}\sum_{q_1=0}^{k_1-1}\sum_{q_2=0}^{k_2-1}\left(-\frac{N}{2}\right)^{2k_1}\frac{(2k_1-q_1-1)!}{q_1!\,(2k_1-2q_1)!}\left(-\frac{N}{2}\right)^{2k_2}\frac{(2k_2-q_2-1)!}{q_2!\,(2k_2-2q_2)!}\nonumber\\
& \left[\langle A_{2k_1-2q_1}^{+}A_{2k_2-2q_2}^{+} \rangle -\langle A_{2k_1-2q_1}^{+} \rangle\langle A_{2k_2-2q_2}^{+} \rangle_0 -\langle A_{2k_2-2q_2}^{+} \rangle\langle A_{2k_1-2q_1}^{+} \rangle_0 + \langle A_{2k_1-2q_1}^{+} \rangle_0 \langle A_{2k_2-2q_2}^{+}\rangle_0
\right]
\label{startPP}
\end{align}
We proceed by following the same method outlined in Appendix B of \cite{Billo:2022xas}. Specifically, we expand the 2-point correlator $\langle A_{2k_1-2q_1}^{+}A_{2k_2-2q_2}^{+} \rangle$ and the 1-point correlator $\langle A_{2k_1-2q_1}^{+}\rangle$ in powers of the interaction action $S_0$. We then explicitly evaluate the first correction in the interaction action. For example, in the case of $\langle A_{2k_1-2q_1}^{+}\rangle$, this amounts to consider
\begin{align}
\langle A_{2k_1-2q_1}^{+}\rangle =  \langle A_{2k_1-2q_1}^{+}\rangle_0 + \langle A_{2k_1-2q_1}^{+}\rangle_0\langle S_0\rangle_0-\langle A_{2k_1-2q_1}S_0 \rangle_0 + O(S_0^2)\, \ , 
\label{expansionA1}
\end{align}
where the interaction $S_0$ has been introduced in \eqref{S0Expansion} and it can be rewritten in terms of the twisted operators $A^{-}_{s}$ as \cite{Billo:2021rdb} 
\begin{align}
S_0 = 2\sum_{m=2}^{\infty}\sum_{k=2}^{2m}(-1)^{m+k}\left(\frac{\lambda}{8\pi^2N}\right)^{m}\left(\begin{array}{c}
     2m  \\
     k 
\end{array}\right)\frac{\zeta_{2m-1}}{m}A_{2m-k}^{-}\,A_{k}^{-}\, \ . 
\label{S0unt}
\end{align}
Therefore, from the expansion \eqref{expansionA1} and the expression \eqref{S0unt}, it becomes evident that we need to derive the large-$N$ expansion of $\langle A_{2s}^{+}A_{n_1}^{-}A_{n_2}^{-}\rangle_0$ and $\langle A_{2s_1}^{+}A_{2s_2}^{+}A_{n_1}^{-}A_{n_2}^{-}\rangle_0$. These are considered in the following section.

\subsection{Large-\texorpdfstring{$N$}{} expansions of \texorpdfstring{$\langle A_{2s}^{+}\,A_{n_1}^{-}\,A_{n_2}^{-} \rangle_0$}{} and \texorpdfstring{$\langle A_{2s_1}^{+}\,A_{2s_2}^{+}\,A_{n_1}^{-}\,A_{n_2}^{-}\rangle_0$}{}}
We use the expressions for the correlation functions in the Gaussian theory and we derive the large-$N$ expansion of
\begin{align}
&\langle A_{2s}^{+}\,A_{n_1}^{-}\,A_{n_2}^{-} \rangle_0 \simeq \nonumber \\
& \langle A_{2s}^{+} \rangle_0\, \langle A_{n_1}^{-}\,A_{n_2}^{-} \rangle_0\,\left[1+\frac{s(s+1)(n_1+n_2)}{4N^2} +  \frac{(n_1+n_2)}{96N^4}g(s,n_1,n_2)+ O(N^{-6}) \right] \,  , 
\label{A1PT}
\end{align}
where $n_1$ and $n_2$ are either odd or even numbers and the function $g(s,n_1,n_2)$ is defined as follows
\begin{align}
s(s+1)\left(p_2(s)\,(n_1^2+n_2^2+n_1n_2)+p_1(s,n_1)\,(n_1+n_2)+p_0(s,n_1)\right) 
\end{align}
and where
\begin{align}
& p_2(s) = (s-1)\, , \ \ \ \ \ \ \  p_1(s,n_1) = 2(s-1)\left(s-8+\frac{3}{2}\delta_{\text{Mod}(n_1,2),1}\right)\,  , \nonumber \\[0.5em]
& p_0(s,n_1) = -s^2(8-6\,\delta_{\text{Mod} (n_1,2),1}) + s(60-63\,\delta_{\text{Mod}(n_1,2),1})-28+57\,\delta_{\text{Mod}(n_1,2),1}\, .
\label{eq:ppolynomials}
\end{align}
We note that the polynomials $p_1$ and $p_0$ exhibit different expressions depending on whether $n_1$ (or equivalently $n_2$) is even or odd. In the same way we have determined the large-$N$ expansion of \footnote{The summation over $i$ in the second line is performed modulo 2, i.e. $i+2\sim i$.}
\begin{align}
& \langle A_{2s_1}^{+}\,A_{2s_2}^{+}\,A_{n_1}^{-}\,A_{n_2}^{-}\rangle \simeq \langle A_{2s_1}^{+}\,A_{2s_2}^{+}\,A_{n_1}^{-}\,A_{n_2}^{-}\rangle_0\,\left[1+\frac{(n_1+n_2)}{4N^2}\sum_{i=1}^2s_i(s_i+1) + \right. \nonumber \\  
& \frac{(n_1+n_2)}{16N^4}\left(\frac{1}{6}\sum_{i=1}^{2}s_i(s_i+1)f(s_i,s_{i+1},n_1,n_2) -\frac{(s_1-s_2)^2}{s_1+s_2}\prod_{i=1}^2s_i(s_i+1)\right) +O(N^{-6}) \Big]\,  , 
\label{A2PT}
\end{align}
where both $n_1$ and $n_2$ are either odd or even numbers. Additionally, the function $f(s_1,s_2,n_1,n_2)$ is defined as follows
\begin{align}
\label{eq:f}
q_2(s_1)\,(n_1^2+n_2^2+n_1\,n_2) + q_1(s_1,s_2,n_1)\,(n_1+n_2)+q_0(s_1,s_2,n_1) 
\end{align}
and where 
\begin{align}
& q_2(s) = (s-1)\, , \nonumber \\[0.5em]
&q_1(s_1,s_2,n_1)=  2s_1^2+ 3s_2^2 -s_1(18-3\,\delta_{\text{Mod} (n_1,2),1})+3s_2+16-3\,
   \delta_{\text{Mod} (n_1,2),1}\, ,\nonumber \\[0.5em]
& q_0(s_1,s_2,n_1) = 6s_2^3 -s_1^2 (8-6 \,\delta_{\text{Mod} (n_1,2),1}) -6s_2^2  
s_1(60-63\,\delta_{\text{Mod} (n_1,2),1}) \nonumber \\
& \qquad \qquad \qquad -12s_2 +57\,\delta_{\text{Mod} (n_1,2),1}-28\, \ .  
\label{eq:qpolynomials}
\end{align}
Similarly to \eqref{eq:ppolynomials}, the expressions for $q_1$ and $q_0$ depend on whether $n_1$ is even or odd.
\subsection{Computation of the first correction}

We are now ready to perform the computation of the terms inside the square bracket of \eqref{startPP}, namely
\begin{align}
\label{Bracket}
& \langle A_{2s_1,2s_2} \rangle - \langle A_{2s_1} \rangle \langle A_{2s_2} \rangle_0 - \langle A_{2s_1} \rangle_0 \langle A_{2s_2}\rangle +2\,t_{2s_1}t_{2s_2}\, \ ,      
\end{align}
where, in order to simplify the notation, we introduced $s_1=k_1-q_1$ and $s_2=k_2-q_2$.
Using the expressions \eqref{A1PT} and \eqref{A2PT}, after a very long but straightforward computation, we derive the large-$N$ expansion of \eqref{Bracket}, which takes the following form
\begin{align}
& (t_{2s_1,2s_2}-t_{2s_1}t_{2s_2})\left[1-\frac{\sum_{i=1}^{2}s_i(s_i+1)}{2N^2}\lambda\partial_{\lambda}\langle S_0 \rangle_0 + O(N^{-4})\right]  -(t_{2s_1,2s_2}+t_{2s_1}t_{2s_2}) \nonumber \\
& \times\left[\frac{\prod_{i=1}^2s_i(s_i+1)}{4N^4}\Big((\lambda\partial_{\lambda})^2\langle S_0 \rangle_0+\frac{2(s_1s_2-s_2-s_1)}{s_1+s_2}\lambda\partial_{\lambda}\langle S_0 \rangle_0 \Big)+ O(N^{-6})\right] + O(S_0^2) \, .
\label{intermediate}
\end{align}
The large-$N$ expansion of the difference $t_{2s_1,2s_2}-t_{2s_1}t_{2s_2}$  reads\,\cite{Billo:2022xas}
\begin{align}
t_{2s_1,2s_2}-t_{2s_1}t_{2s_2} \,\simeq \,T^{\text{LO}}_{s_1,s_2}\,\alpha_{s_1}\alpha_{s_2} + T^{\text{NLO}}_{s_1,s_2}\,\alpha_{s_1}\alpha_{s_2} + \dots     
\end{align}
where the dots stand for subleading terms of the expansion while the other quantities are defined as
\begin{align}
& T_{s_1,s_2}^{\text{LO}} = \frac{1}{s_1+s_2}\, , \quad T_{s_1,s_2}^{\text{NLO}} = \frac{1}{12N^2}(s_1^2+s_2^2+s_1\,s_2-8(s_1+s_2)+7)\, ,\quad  \alpha_{s_1} = \frac{N^{s_1}(2k_1-1)!!}{(s_1-1)!}\, .
\end{align}
While the large-$N$ expansion of the sum $t_{2s_1,2s_2}+t_{2s_1,2s_2}$ is given by
\begin{align}
t_{2s_1,2s_2}+t_{2s_1,2s_2} \simeq  \frac{2}{\prod_{i=1}^{2}s_i(s_i+1)}\alpha_{s_1}\alpha_{s_2}+ \dots   
\end{align}
where again the dots denote subleading terms.
Hence, assuming that the same pattern persists when considering higher powers of $S_0$, we can substitute $\langle S_0 \rangle_0$ in \eqref{intermediate} with the free energy $\mathcal{F}$. Therefore, we are only left with evaluating the finite sums over $q_1$ and $q_2$ in \eqref{startPP}. These can be computed analytically by exploting the following identities
\begin{subequations}
\begin{align}
& \sum_{q=0}^{k-1}\frac{(-1)^q\,(2k-q-1)!}{(k-q-1)!\,(k-q)!\,q!} = 1\, , \\[0.5em]
& \sum_{q=0}^{k-1}\frac{(-1)^{q}\,(2k-q-1)!\,q}{(k-q-1)!\,(k-q)!\,q!} = -k(k-1)\,  , \\[0.5em]
& \sum_{q=0}^{k-1}\frac{(-1)^{q}\,(2k-q-1)!\,q^2}{(k-q-1)!\,(k-q)!\,q!} = \frac{k^{2}(k-1)(k-3)}{2}\, , \\[0.5em]
& \sum_{q_1=0}^{k_1-1}\sum_{q_2=0}^{k_2-1}\left(\frac{(-1)^{q_1}(2k_1-q_1-1)!}{(k_1-q_1-1)!\,(k_1-q_1)!\,q_1!}\,\frac{(-1)^{q_2}(2k_2-q_2-1)!}{(k_2-q_2-1)!\,(k_2-q_2)!\,q_2!}\frac{1}{k_1+k_2-q_1-q_2} \right) = \frac{\delta_{k_1,k_2}}{2k_1}\, .
\end{align}
\end{subequations}
After a long computation we analytically obtain
\begin{align}
\langle & \mathcal{P}_{2k_1}^{+}\mathcal{P}_{2k_2}^{+} \rangle \simeq \nonumber \\
&\delta_{k_1,k_2}+\frac{\sqrt{k_1\,k_2}}{N^2}\left[\frac{(k_1^2+k_2^2-1)(k_1^2+k_2^2-14)}{12} - (k_1^2+k_2^2-1)\lambda \partial_{\lambda}\mathcal{F} -(\lambda\partial_{\lambda})^2\mathcal{F} +O(N^{-4}) \right]\, .
\end{align}
We notice that the expression above differs from \eqref{Tplus} by the additional term $\sqrt{k_1k_2}\,(\lambda\partial_{\lambda}\mathcal{F})^2$. However, this discrepancy is expected since the analysis conducted in the appendix has been limited to the first correction in the interaction action, while the omitted contribution can only be accounted for by terms quadratic in $\langle S_0 \rangle_0^2$. While it would certainly be very interesting to derive explicitly the second correction in the interaction action, such a computation goes beyond the scope of this paper. Nevertheless, we would like to point out that we have thoroughly verified, perturbatively, by expanding up to very high orders in the 't Hooft coupling, the validity of the relation \eqref{Tplus}.

\printbibliography

\end{document}